\documentclass[12pt,preprint]{aastex}
\usepackage{psfig,natbib}
\shortauthors{Bower et al.}
\shorttitle{PiGSS I}
\begin{document}

\newcommand\degd{\ifmmode^{\circ}\!\!\!.\,\else$^{\circ}\!\!\!.\,$\fi}
\newcommand{\etal}{{\it et al.\ }}
\newcommand{\uv}{(u,v)}
\newcommand{\rdm}{{\rm\ rad\ m^{-2}}}
\newcommand{\msuny}{{\rm\ M_{\sun}\ y^{-1}}}
\newcommand{\mylesssim}{\stackrel{\scriptstyle <}{\scriptstyle \sim}}
\newcommand{\lsim}{\stackrel{\scriptstyle <}{\scriptstyle \sim}}
\newcommand{\gsim}{\stackrel{\scriptstyle >}{\scriptstyle \sim}}
\newcommand{\sci}{Science}


\title{The Allen Telescope Array Pi GHz Sky Survey I.  
Survey Description and Static Catalog Results for the Bo\"{o}tes Field}

\author{
Geoffrey C.\ Bower\altaffilmark{1},
 Steve Croft\altaffilmark{1}, 
 Garrett Keating\altaffilmark{1},
 David Whysong\altaffilmark{1},
 Rob Ackermann\altaffilmark{2},
 Shannon Atkinson\altaffilmark{2},
 Don Backer\altaffilmark{1},
 Peter Backus\altaffilmark{2},
 Billy Barott\altaffilmark{3},
 Amber Bauermeister\altaffilmark{1},
 Leo Blitz\altaffilmark{1},
 Douglas Bock\altaffilmark{1},
 Tucker Bradford\altaffilmark{2},
 Calvin Cheng\altaffilmark{1},
 Chris Cork\altaffilmark{4},
 Mike Davis\altaffilmark{2},
 Dave DeBoer\altaffilmark{5},
 Matt Dexter\altaffilmark{1},
 John Dreher\altaffilmark{2},
 Greg Engargiola\altaffilmark{1},
 Ed Fields\altaffilmark{1},
 Matt Fleming\altaffilmark{4},
 R. James Forster\altaffilmark{1},
 Colby Gutierrez-Kraybill\altaffilmark{1},
 G.R. Harp\altaffilmark{2},
 Carl Heiles\altaffilmark{1},
 Tamara Helfer\altaffilmark{1},
 Chat Hull\altaffilmark{1},
 Jane Jordan\altaffilmark{2},
 Susanne Jorgensen\altaffilmark{1},
 Tom Kilsdonk\altaffilmark{2},
 Casey Law\altaffilmark{1},
 Joeri van Leeuwen\altaffilmark{6},
 John Lugten\altaffilmark{7},
 Dave MacMahon\altaffilmark{1},
 Peter McMahon\altaffilmark{8},
 Oren Milgrome\altaffilmark{1},
 Tom Pierson\altaffilmark{2},
 Karen Randall\altaffilmark{2},
 John Ross\altaffilmark{2},
 Seth Shostak\altaffilmark{2},
 Andrew Siemion\altaffilmark{1},
 Ken Smolek\altaffilmark{2},
 Jill Tarter\altaffilmark{2},
 Douglas Thornton\altaffilmark{1},
 Lynn Urry\altaffilmark{1},
 Artyom Vitouchkine\altaffilmark{4},
 Niklas Wadefalk\altaffilmark{9},
 Sandy Weinreb,
 Jack Welch\altaffilmark{1},
 Dan Werthimer\altaffilmark{1},
 Peter K.G. Williams\altaffilmark{1}, and
 Melvyn Wright\altaffilmark{1}}
\altaffiltext{1}{University of California, Berkeley, 601 Campbell Hall \#3411, Berkeley, CA 94720, USA; gbower@astro.berkeley.edu }
\altaffiltext{2}{SETI Institute, Mountain View, CA 94043, USA }
\altaffiltext{3}{Embry-Riddle Aeronautical University, Electrical and Systems Engineering Department, Daytona Beach, FL 32114, USA }
\altaffiltext{4}{Minex Engineering, Antioch, CA 94509, USA }
\altaffiltext{5}{CSIRO/ATNF, Epping, NSW 1710, Australia }
\altaffiltext{6}{ASTRON, 7990 AA Dwingeloo, The Netherlands}
\altaffiltext{7}{Lawrence Livermore National Laboratory, Livermore, CA 94550, USA}
\altaffiltext{8}{Electrical Engineering Department, Stanford University, Stanford, CA 94305, USA }
\altaffiltext{9}{Chalmers University of Technology, Department of Microtechnology and Nanoscience - MC2, SE-412 96 G\"{o}teborg, Sweden}
\altaffiltext{10}{California Institute of Technology, Department of Electrical Engineering, Pasadena, CA 91125, USA}

\begin{abstract}
The Pi GHz Sky Survey (PiGSS) is a key project of the Allen Telescope Array.  PiGSS
is a 3.1 GHz survey of radio continuum emission in the extragalactic sky with an
emphasis on synoptic observations that measure the static and
time-variable properties of the sky.  During the 2.5-year campaign, PiGSS will
twice observe $\sim 250,000$ radio sources in the 10,000 deg$^2$ region of the sky with
$b > 30^\circ$ to an rms sensitivity of $\sim 1$ mJy.  Additionally, sub-regions of the 
sky will be observed multiple times to characterize variability on time scales of
days to years.  We present here observations of a 10 deg$^2$ region in the Bo\"{o}tes
constellation overlapping the NOAO Deep Wide Field Survey field.  
The PiGSS image was constructed from 75 daily observations distributed over a
4-month period and has an rms flux density between 200 and 
250 $\mu$Jy.  This represents a deeper image by a factor of 4 to 8 than 
we will achieve over the entire 10,000 deg$^2$.  We provide flux densities, source sizes,
and spectral indices for the 425 sources detected in the 
image.  We identify $\sim 100$ new flat spectrum radio sources; we 
project that when completed PiGSS will identify $10^4$ flat spectrum sources.
We identify one source that is a possible transient radio source.  This survey provides 
new limits on faint radio transients and variables with characteristic durations
of months.  
\end{abstract}

\keywords{radio continuum:  general --- radio continuum:  stars ---
radio continuum:  galaxies --- surveys}

\section{Introduction}

Synoptic surveys are increasingly important probes of the sky at all wavelengths.  
These surveys provide the opportunity to simultaneously probe the static and
variable components of the Universe through multiple observations of the sky.
Complete static catalogs such as that of the NRAO VLA Sky Survey 
\citep{1998AJ....115.1693C},
the VLA FIRST Survey
\citep{1995ApJ...450..559B}, the Sloan Digital Sky Survey
\citep{2003AJ....126.2081A}, and the ROSAT
All Sky Survey 
\citep{1999A&A...349..389V}
are essential to creating a complete picture of the astrophysics of a
wide range of objects.  These extant surveys predominantly emphasized
a single static image of the sky.  The next generation of large-scale surveys
at all wavelengths have variability and the time domain as an essential aspect.
These surveys include at optical wavelengths
the SDSS stripe 82 survey 
\citep{2007AJ....134.2236S},
the Palomar Transient Factory 
\citep{2009PASP..121.1334R,2009PASP..121.1395L}, 
Pan-STARRS,
the LSST surveys,
and at gamma-ray wavelengths the Fermi LAT catalog \citep{2010ApJS..188..405A}.

At radio wavelengths, time domain surveys have a venerable history 
including the discovery of pulsars  
\citep{1969Natur.224..472H} but are undergoing a significant 
renaissance.  In particular, single dish surveys with sensitivity to
short timescale transients ($\lsim 1$ second) have uncovered a wide
range of neutron star phenomena, including single pulses detected from
rotating radio transients \citep{2006Nature...mcl}.  These discoveries
have spawned several new surveys 
\citep{2008arXiv0811.3046S}.
Future versions 
of single-pulse surveys
are proposed for several next generation instruments such as
ASKAP, LOFAR, MEERKAT, and the Square Kilometer Array
\citep{2010arXiv1001.2958M,2009ASPC..407..318H}.  The enormous data volumes of 
these surveys require novel and
dedicated signal processing techniques.  

Imaging synoptic surveys, however,
are in their infancy.  They have been conducted systematically over
limited area on the sky or assembled from archival data.
Nevertheless, these limited surveys indicate that there are many phenomena
to explore.  VLA surveys of the Galactic Center at low
frequency, for example, have
uncovered several transients without clear identification 
\citep{2002AJ....123.1497H,2005Natur.434...50H,2009ApJ...696..280H}.
A comparison of the FIRST and NVSS surveys also uncovered a radio
supernova and a radio transient without an optical counterpart 
\citep{2002ApJ...576..923L,2006ApJ...639..331G,2010ApJ...711..517O}.
\citet{2007ApJ...666..346B}
discovered 10 radio transients without 
counterparts in deep optical and radio images from nearly 1000 observations
of a single field obtained weekly over 20 years.  
Recently, \citet{2010AJ....140..157B} discovered a number of variable
radio sources in observations of the Galactic Plane.
In addition to
objects and events discovered through systematic searches, radio
transients and variables have frequently been found through
serendipity.  
\citet{2009A&A...499L..17B},
for instance, discovered
a radio supernova in M82 during VLBI observations intended to
characterize the proper motion of the galaxy.
The time scale of days to years includes known radio variability from
radio supernovae and
gamma-ray burst afterglows 
\citep{2002ARA&A..40..387W}, interstellar propagation such as extreme-scattering
events 
\citep[ESEs;][]{1987ApJS...65..319F,2008ApJ...672L..95S}, intrinsic AGN processes 
\citep{1992ApJ...396..469H}, and compact-object binaries and stars 
\citep[e.g.,][]{2001Natur.410..338B,2005ApJ...633..218B,Osten06}.

\citet{2010A&ARv..18....1D} summarize radio continuum surveys and 
their astrophysical uses.  
The NRAO VLA Sky Survey \citep[NVSS;][]{1998AJ....115.1693C}
and the FIRST survey \citep{1995ApJ...450..559B}
constitute the best existing 1.4 GHz radio surveys of the sky.
The best existing high frequency survey is the GB6 catalog, which has a detection
threshold of $\sim 25$ mJy \citep{1996ApJS..103..427G}.  
Large area, high frequency surveys that can 
provide statistical information about spectral indices and sources 
populations include the
AT20G survey, which covers the southern hemisphere
to a flux density threshold of 100 mJy (91\% completeness) at 20 GHz \citep{2010MNRAS.402.2403M},
and the 9C survey, which covers 520 square degrees to 25 mJy at 15 GHz 
\citep{2010MNRAS.404.1005W}.

The static component of large scale surveys is valuable for a wide
range of astrophysical problems.  New radio surveys have the capability
of providing spectral index information for radio sources, which can
be critical for separating source classes.  
This includes separation of star-forming galaxies
from active galactic nuclei \citep{2002AJ....124.2364I}, identification of GHz-spectrum-peaked sources \citep{1998A&AS..131..303S},
and identification of flat spectrum radio sources.  
In particular, flat spectrum catalogs are used for identification of 
blazars that may be gamma-ray counterparts \citep{2008ApJS..175...97H,2009A&A...495..691M} 
and that may be
useful for flux-monitoring in determination of gravitational lensing
constraints on cosmological parameters \citep{2003MNRAS.341....1M,2002ApJ...581..823F}.

We describe here a new synoptic survey at radio wavelengths
conducted with the Allen Telescope Array (ATA)
and provide initial results.  
The Pi GHz Sky Survey (PiGSS) is a key project of the ATA.  
Key goals are 1) to conduct a large-area survey of radio
continuum at a frequency of 3.1 GHz that approaches an order of magnitude more
sensitivity 
than the best existing catalog at frequencies higher than 1.4 GHz; and 2) to explore
the variable and transient component of the radio sky with a method
that is unbiased by optical or high energy observations.  The first goal will produce
spectral indices for an order of magnitude more sources than currently exist,
enabling a wide range of science.  This survey builds on a preliminary
ATA project, the ATA Twenty Centimeter Survey \citep[ATATS;][]{2010ApJ...719...45C},
which also demonstrates imaging quality and array performance.
We describe the ATA, the survey, and its goals in \S 2 and the data
reduction techniques in \S 3.  We present our results in \S 4 and
discuss these in \S 5.  We summarize in \S 6.

\section{Survey Description}

The Allen Telescope Array (ATA) is a new radio telescope designed to conduct surveys 
\citep{2009arXiv0904.0762W}.
The ATA is a pioneer of the large-N-small-diameter (LNSD) array design, which
characterizes the international approach to the Square Kilometer Array design
\citep{schilizzi_100}.
The survey speed is the rate per unit time
at which a telescope covers solid angle to a fixed
sensitivity.  For a radio interferometer with $N$
elements of diameter $D$ the survey speed scales
as $(ND)^2$, whereas the time to observe a single field scales proportional to 
the square of the total area $(ND^2)^2$.  Thus, for a fixed collecting area, a
smaller diameter $D$ will provide faster survey speed.
Unique features of the ATA are a wide field of view ($2.5^\circ$ at 1.4 GHz),
compact configuration ($b_{max} = 300$ meters), a broadband feed
that delivers the entire radio frequency band of 0.5 to 11 GHz in orthogonal
linear polarizations to the
laboratory, and flexible digital signal processing that includes two correlators
and three phased-array beamformers and uses four individual frequency tunings.
Each correlator has a bandwidth of 104 MHz in 1024 channels and provides
full Stokes information.
\citet{2010ApJ...710.1462W}
have exploited the broad frequency coverage of the telescope to measure continuous
spectra of flux calibration standard sources and nearby star-forming galaxies,
including M82, NGC 253, and Arp 220.  The ATA was dedicated in Fall 2007; commissioning
observations have been carried out over the past 2.5 years as array performance
has improved.

The observing frequency was selected by balancing several factors:  the need
for higher frequency to achieve sufficient leverage to measure spectral indices;
the expectation of greater amplitude and more rapid timescale variability from 
synchrotron sources at higher frequencies; the decreasing field of view and, hence,
survey speed with increasing frequency; the improving sensitivity of the ATA at 
lower frequencies; and the absence of significant radio frequency interference (RFI).
Two adjacent frequency bands were selected for the two correlators with center
frequencies of 3.04 and 3.14 GHz.

Field selection for PiGSS was based on the goal of covering a very large area of
extragalactic sky and overlapping with existing deep and wide area surveys.  
An overview of fields and specifications is given in Table~\ref{tab:pigss}.
The
North Galactic Cap at $b > 30^\circ$ has been extensively imaged, in particular by
the Sloan Digital Sky Survey 
\citep{2003AJ....126.2081A} at optical wavelengths and by 
the NVSS and FIRST surveys
at a frequency of 1.4 GHz.  Individual sub-fields were selected for similar reasons.
The NOAO Deep Wide Field Survey (also known as the Bo\"{o}tes field)
\citep{2004AAS...205.8106J,2009ApJ...701..428A} has been extensively imaged from radio to X-ray
wavelengths to significant depth and over an area as large as 10 deg$^2$.  PiGSS
is also targeting a 10 deg$^2$ region covering the Lockman hole for similar
reasons.  Both fields are seven-pointing hexagonally-close-packed
mosaics with a spacing between pointings equal to the voltage full width at half
maximum, $\Delta\theta=0.78^\circ$ at 3.14 GHz.  Additional 10 deg$^2$ fields
will be selected and observed throughout the survey.  A 250 deg$^2$ field overlapping
the Bo\"{o}tes field was also observed and will be the subject of a subsequent paper.

The tiling of the North Galactic Pole (NGP) mosaic attempts to achieve a hexagonally-close-packed spacing
locally while minimizing the total number of pointings.  Rows of pointings at 
constant declination are arranged with optimal spacing in right ascension 
for the minimum declination
within that band of rows; i.e., $\Delta\alpha_i = \Delta\theta / \cos (\delta_{i,min})$,
where $\delta_{i,min}$ is the minimum declination within a band of rows.  
Pointing centers in adjacent rows are offset in right ascension
by half of the spacing to provide the close-packed spacing.  The number of rows
for a fixed $\Delta\alpha_i$ is set by the constraint that the number of pointings
per row does not exeed 110\% of the optimal number for that row.  The total number
of pointings for the NGP is 19940.

The integration time was selected
with the goal of ultimately achieving a $5\sigma$ detection threshold of 5 mJy for a single
epoch and 4 mJy for the completed large area survey.
Typical observations consist of a loop that involves two minute observations of
each pointing in one of the 10 deg$^2$ fields and approximately 50 pointings
in the NGP (or other large) field.  These observations are repeated between three
and six times in an observing session with a duration of between 6 and 18 hours.  
This repetition gives more uniform $(u,v)$ coverage over the fields as well as increasing
sensitivity.  As the number of antennas available to the correlator increases, the
number of observations per field wlll be reduced without losing sensitivity and 
$(u,v)$ coverage.  Commissioning of new antennas and of new correlator capacity has
been proceeding along with this survey.  The observations reported here rely on data
from about 25 dual polarization antennas.

The longest baseline of the ATA currently is $\sim$300 meters, providing 
$2\arcmin \times 1\arcmin$ resolution at 3.1 GHz .  This resolution is a good match to
NVSS ($\sim$45$^{\prime\prime}$), enabling accurate spectral indices.
Source positions will be accurate to
$10^{\prime\prime}$ or better.  Matching with known FIRST sources will permit
$1^{\prime\prime}$ accuracy positions for the majority of sources.

Source confusion from integrated faint sources provides a limit on the ultimate
sensitivity that a static survey can obtain.  Transient and variable surveys
can, of course, detect variable sources that are fainter than the confusion 
limit.  \citep{1974ApJ...188..279C}
derived the standard equations for confusion noise
due to faint background sources.  The results depend solely on the
telescope resultion and the 
differential number count of sources per unit solid angle
\begin{equation}
n(S) dS = k S^{-\gamma} dS,
\end{equation}
where $k$ is a normalization constant and $\gamma \approx 2$ is the power-law
index.
The confusion noise is
\begin{equation}
\sigma_c = \left( q^{3-\gamma} \over 3 - \gamma \right)^{1 \over \gamma -1}
           \left( k \Omega_e\right)^{1 \over \gamma -1}.
\end{equation}
$\Omega_e$ is the effective beam area 
\begin{equation}
\Omega_e = {\pi \over 4} \theta_1 \theta_2 / \left[ \left( \gamma - 1 \right) \ln 2 \right],
\end{equation}
where $\theta_1$ and $\theta_2$ are the major and minor FWHM axes of 
the telescope beam.  $q$ is an arbitrary parameter of the
integration which is typically set to 5.
The detection threshold of an array is $\sim 5$ times the confusion
noise for a nonvariable source.
For the ATA, the rms confusion at 1.4 GHz and 5.0 GHz as 325 $\mu$Jy and 
40 $\mu$Jy, respectively, based on source counts at those wavelengths.  
In the absence of detailed source counts at 3.1 GHz, we can interpolate
between those results and estimate
the rms confusion at 3.1 GHz to be $\sim$ 150 $\mu$Jy.

Cadence for synoptic observations is set to satisfy scheduling constraints and 
provide sensitivity to transient and variable sources on time scales of
days to months.  For the
10,000 deg$^2$ field, observations will be performed yearly, providing two samples
of the flux density.  For the 10 deg$^2$ fields, a cadence with intervals of
days to years will be obtained.  Following 
\citet{2007ApJ...666..346B},
the two-epoch rate for
a transient survey with $N_e$ epochs that cover an area $A$ to sensitivity $S$ is 
\begin{equation}
R(>S)={N_t \over {(N_e-1) A(>S)}}\ ,
\label{eqn:snapshot}
\end{equation}
where $N_t$ is the number of transients detected.  Here, $A(>S)$ refers to 
the solid angle over which a source of flux density $S$ or greater can 
be detected.  Where no transient is 
detected, the $2\sigma$ limit is $N_t \approx 3$
\citep{1986ApJ...303..336G}.
\citet{2010ApJ...719...45C}
provide an updated version of this two-epoch rate.  This rate applies to 
transients with a timescale comparable to or longer than the duration of individual epochs.
Thus, the optimal approach to maximize the sensitivity to transients is to
repeatedly  
observe the same field.  Comparison to a prior epoch (such as NVSS or GB6) 
can increase $N_e$ by 1.


\section{Observing \& Data Reduction}

\begin{figure}[tb]
\psfig{figure=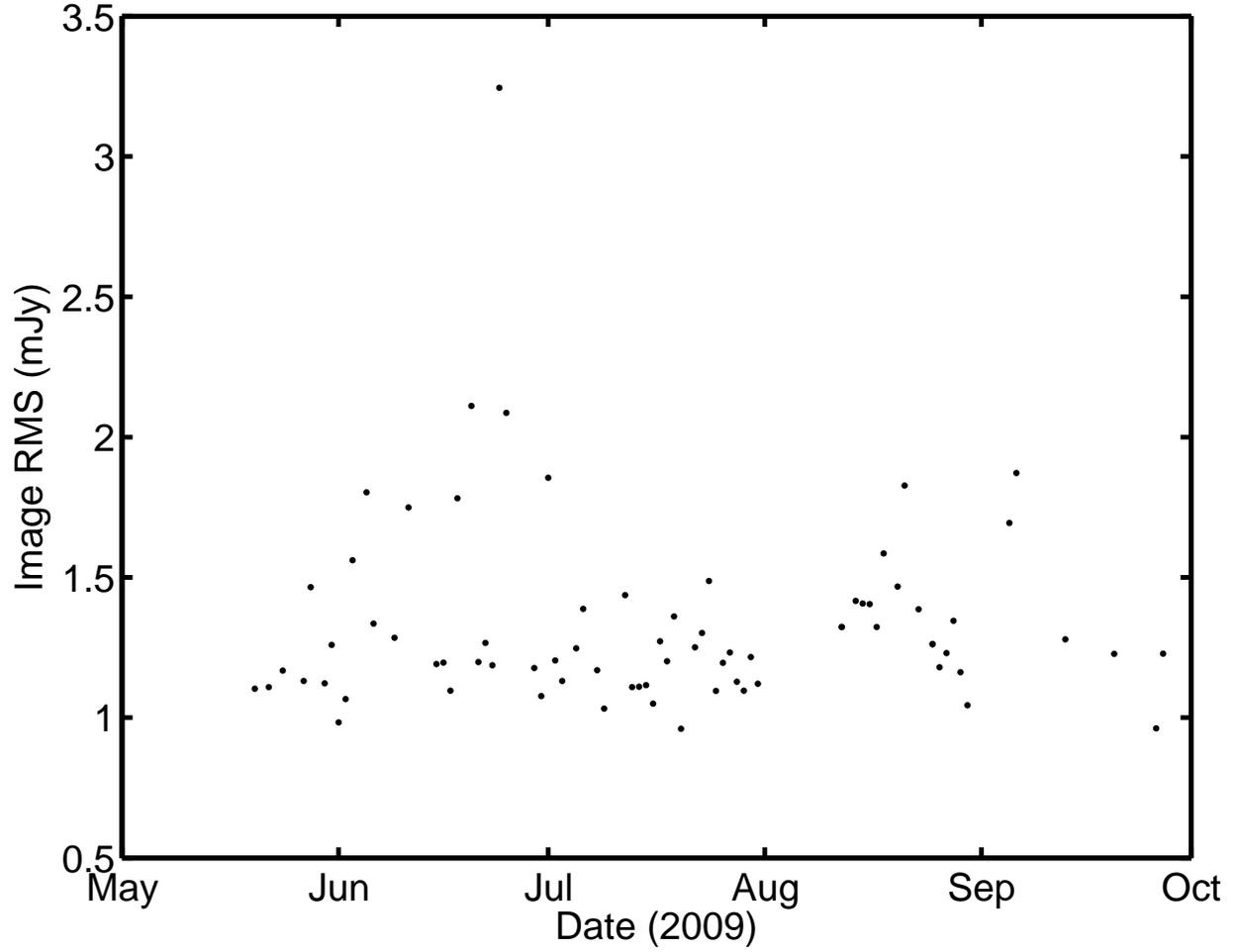,width=\textwidth}
\caption[]{RMS for each of the Bo\"{o}tes epochs as a function of time.
\label{fig:rms}
}
\end{figure}

PiGSS observing began in May 2009.  We report here on observations between 2009 May 20
and 2009 September 27.  A total of 96 observations of the Bo\"{o}tes field
were obtained, of which 75 were included in the final result.  Observations in 
early August 2009 were interrupted for a period of two weeks due to forest fires
in the region; observations in September 2009 were scheduled less densely than earlier
in the campaign.  We show the image rms of epochs with good data in Fig.~\ref{fig:rms}.

\subsection{Calibration and Flagging}

3C 286 was used as the calibrator for the amplitude scale as well as for short-term
variations in antenna amplitude and phase gains.  Calibrator observations
were obtained hourly.  A flux density
of 9.7 Jy was used, following the model of 
\citet{1977A&A....61...99B}.
Observations of the bright 
calibrator 3C 295 on 26 September 2009 
indicate that the amplitude scale is accurate at the level of  $\sim 1\%$,
consistent with earlier measurements by 
\citet{2010ApJ...710.1462W}
who find an
accuracy of $\sim 3\%$.  Total flux density in each of the two correlator bands
was measured to be $10.84 \pm 0.02$  Jy and $10.55 \pm 0.05$ Jy; the expected flux
density at these frequencies is 10.845 and 10.493 Jy, respectively.

Data were reduced using the ARTIS package of scripts for RFI 
excision and amplitude calibration 
\citep{2009AAS...21460106K}.
ARTIS uses the MIRIAD package 
\citep{1995adass...4..433S}.
Some additional RFI excision and rejection of bad antennas and
baselines was performed by hand.  The pipeline currently does not perform
polarization leakage calibration; therefore, we do not provide polarimetric
results in this paper.  Accurate polarization calibration with the ATA 
has recently been demonstrated 
\citep{law_paper}; we anticipate that polarization calibration
will be included in future versions of the pipeline.  The lack of
polarization calibration probably introduces a $\sim 1$\% error in total
flux density and may limit dynamic range
\citep{2010ApJ...719...45C}.

\subsection{Imaging}

All baselines shorter than $80 \lambda$ 
were rejected to exclude large-scale structure, solar interference, and
cross-talk from closely-spaced antennas from the resulting image.
This has the effect of removing any sources with an angular size larger
than $\sim 1^\circ$.
Multi-frequency synthesis imaging was performed independently
for each of the two frequency bands.  A typical synthesized beam for PiGSS 
had a size of $120\arcsec \times 60\arcsec$ with a position angle that varied
depending on the hour-angle range observed; we restored the image with
a synthesized beam of $100\arcsec \times 100\arcsec$.
A linear mosaic of all 7 pointings at
a single frequency was obtained using the primary beam
tapering algorithm described in
\citet{1996A&AS..120..375S}.  The tapered map provides uniform noise across
the image, which is useful for automated source identification, at the expense
of uniform gain.  We apply an after the fact amplitude calibration on 
sources by dividing fluxes by the tapering gain.
The mosaics at the different frequencies
were then combined with
a weighted sum based on the measured noise in the maps.  Finally, the mosaicked images
were converted into the GLS geometry, which provides pixels of uniform area over
the full field \citep{2002A&A...395.1077C}.  Without this correction, fluxes determined
by our source detection algorithm have errors that increase with distance from the
map center.  We plot
the rms of the final images from each epoch in Fig.~\ref{fig:rms}.  The median rms for the
set of images is 1.2 mJy.

The measured primary beam FWHM of the ATA is $3.5^\circ f^{-1}$, where $f$ is the frequency in GHz
\citep{2009arXiv0904.0762W}.
We confirm through examination of sources that are detected
in two or more pointings that the best fit primary beam is consistent with
the nominal value of 1.13$^\circ$ for 3.1 GHz 
\citep{hull_paper}.
Mosaics are imaged to 5\% of the beam sensitivity for a total area of 11.7 deg$^2$.

\begin{figure}[tb]
\psfig{figure=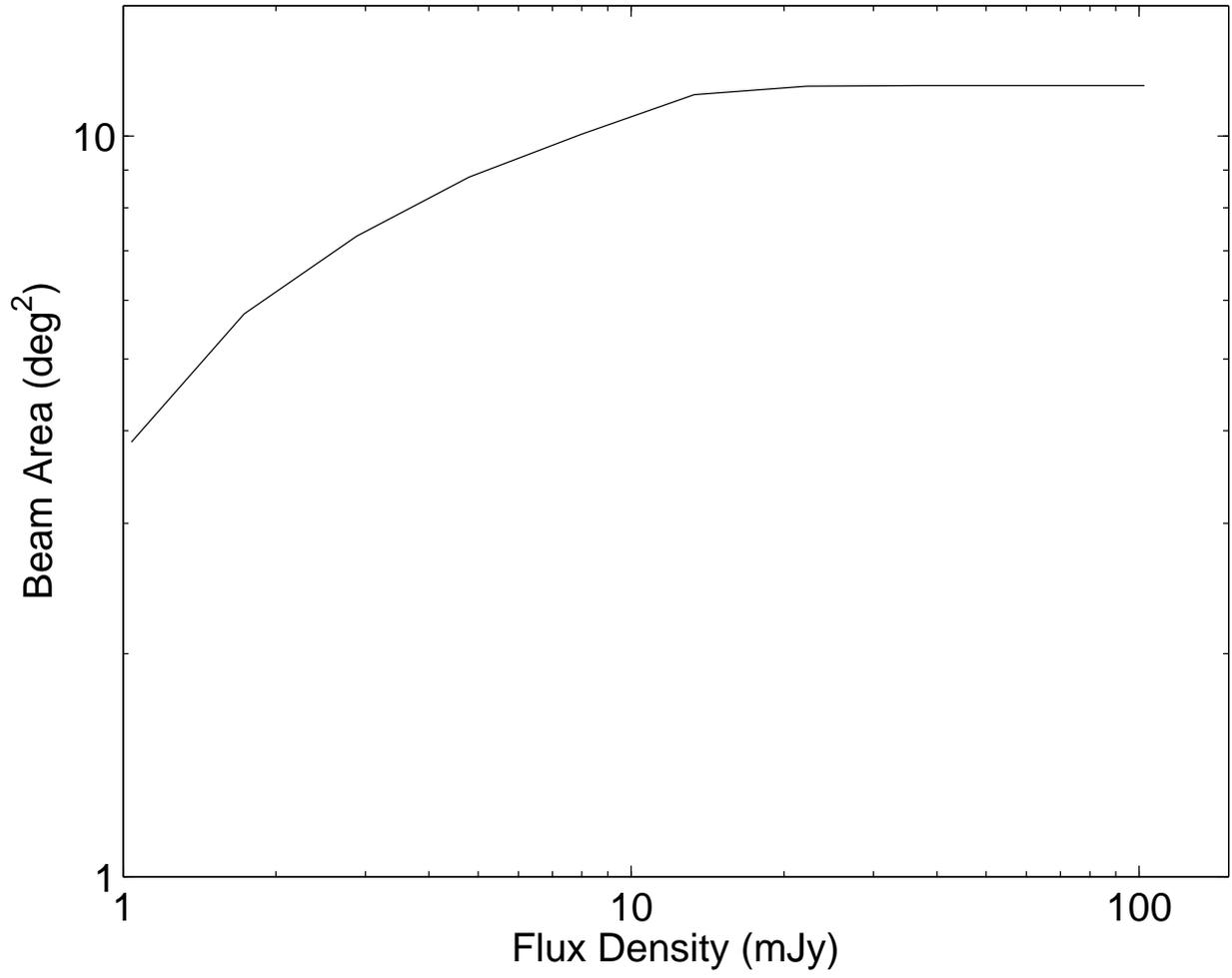,width=\textwidth}
\caption[]{Imaged area as a function of flux density threshold.
\label{fig:beamarea}
}
\end{figure}

\begin{figure}[tb]
\psfig{figure=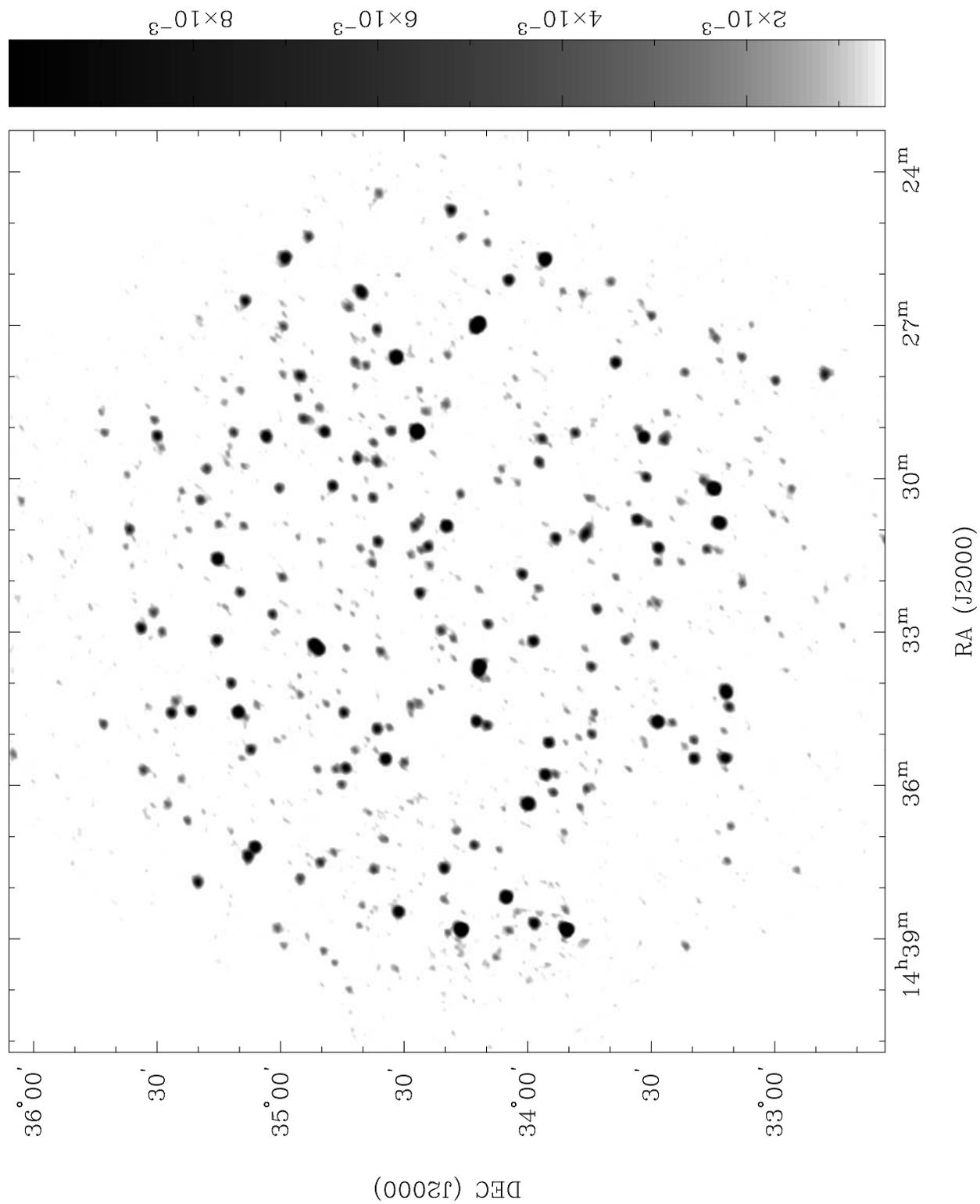,width=\textwidth}
\caption[]{Image of the Bo\"{o}tes field at 3.1 GHz.  The wedge demonstrates the
flux range of the grayscale.
\label{fig:deep}
}
\end{figure}

A deep image of the Bo\"{o}tes field was obtained by merging calibrated
visibility data for each pointing for the entire observation period.  
This merged
data set was imaged and then all pointings were combined to produce a linear
mosaic of the field. 
This image (Fig.~\ref{fig:deep}) has an rms
flux density between 216 and 250 $\mu$Jy, as measured in regions free of sources.
The residual image after subtraction of CLEAN components has an rms of 200 $\mu$Jy.
If we sum in quadrature the rms noise from each of the daily images,
we obtain an expected rms noise of 145 $\mu$Jy.  Additional noise in the image 
may be due to residual RFI, miscalibrated antennas, incomplete cleaning, and
source confusion.  The image includes data from approximately 150 hours of 
observing using approximately half of the array.  With the full array, the same
image can be made in about 40 hours of observing.
The total area with sensitivity above 25\% of the peak 
sensitivity ($\sim 1$ mJy rms) is 5.5 deg$^2$;  above 50\% of the peak sensitivity 
($\sim 0.5$ mJy rms) the total area is
3.3 deg$^2$ (Fig.~\ref{fig:beamarea}).

\subsection{Source Catalog Construction}

We constructed a catalog of sources using the MIRIAD program {\tt sfind} 
\citep{2002AJ....123.1086H}.  We constrain source-fitting to be an elliptical
Gaussian with a minimum size of the synthesized beam.  This has the effect
that many point sources will have an exact deconvolved size and position
angle of 0.
We reject a small number of sources for which Gaussian fitting does not provide an adequate 
solution; these sources appear spurious (typically very elongated) 
when examined in the image domain.  Source
positions ($\alpha, \delta$ in equinox J2000), flux densities, measured Gaussian sizes
($b_{maj}, b_{min}, \phi$), and 
deconvolved Gaussian sizes 
($b_{maj}^\prime, b_{min}^\prime, \phi^\prime$)
are tabulated in Table~\ref{tab:sources}.  
We determine the completeness of the catalog through comparison with the NVSS
catalog, as discussed in the next section.

\section{Bo\"{o}tes Field Results}

For the Bo\"{o}tes field, 
we apply a second cut to the catalog of a minimum flux density of 1 mJy, which is 4 to 5  times the
rms noise in the residual image.  The total number of sources 
is then 425.  
The PiGSS catalog contains more than 10 times as many as sources as the 
GB6 catalog, which has 37 sources over this region.  The
minimum PiGSS flux is a factor of 2.5 and 18 fainter than the minimum NVSS
and GB6 fluxes, respectively.

The Bo\"{o}tes field has been investigated systematically across the spectrum.  Critical radio
surveys against which
to compare PiGSS data are the all sky NVSS and GB6 surveys and a 
deep Westerbork survey at 1400 MHz \citep{2002AJ....123.1784D}. 
We summarize the characteristics of these 
surveys in Table~\ref{tab:bootes}.  We focus on comparison with NVSS and GB6 because these
represent large-area surveys at resolutions similar to PiGSS.  Comparison of the FIRST survey
directly to PiGSS is likely to introduce source mismatches as sources are resolved into
multiple components.  Comparisons of FIRST and NVSS catalogs have been extensively 
investigated \citep[e.g.,][]{2006ApJ...639..331G}.

\begin{figure}[tb]
\mbox{\psfig{figure=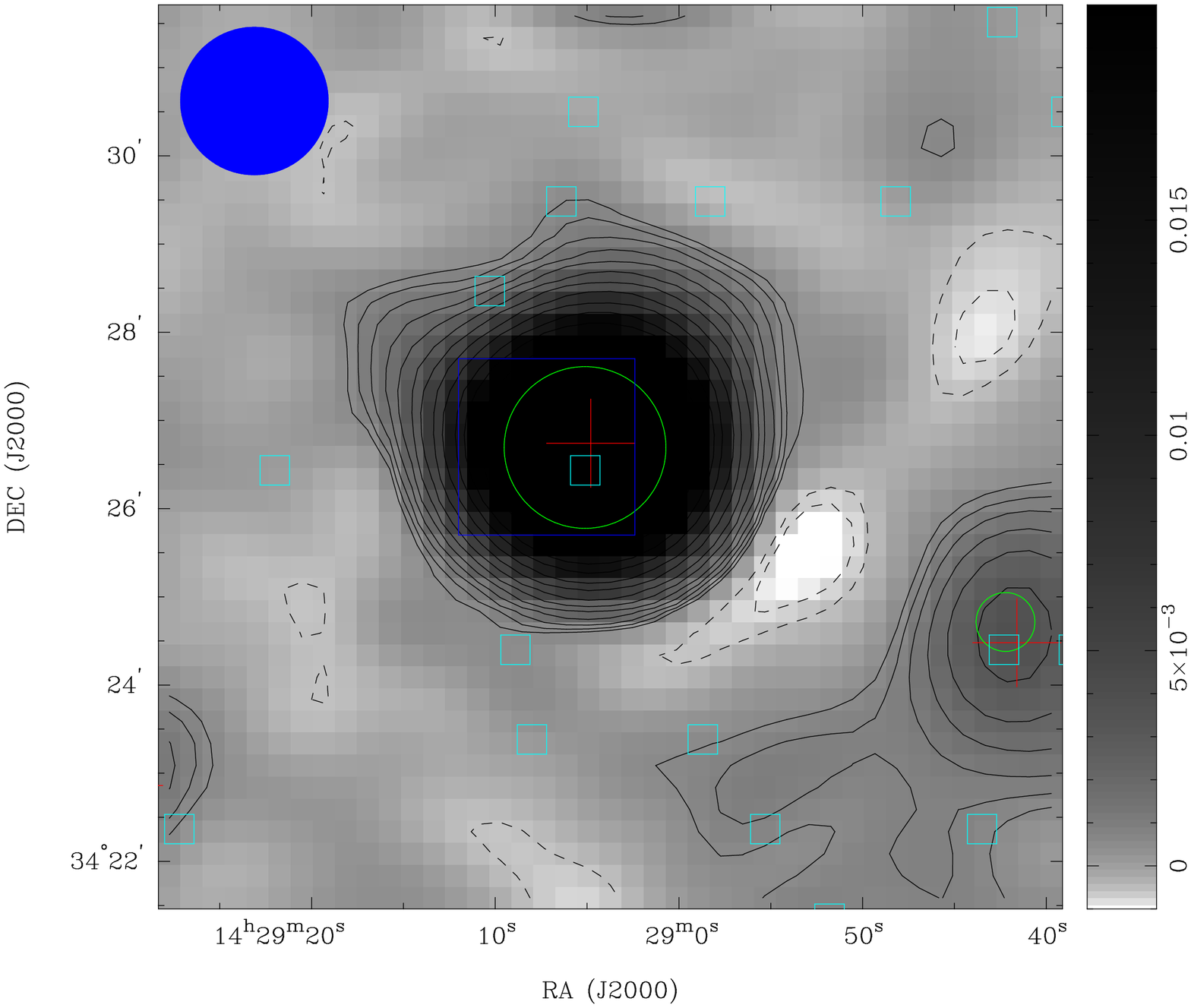,width=0.42\textwidth,angle=-90}\hspace{0.1\textwidth}\psfig{figure=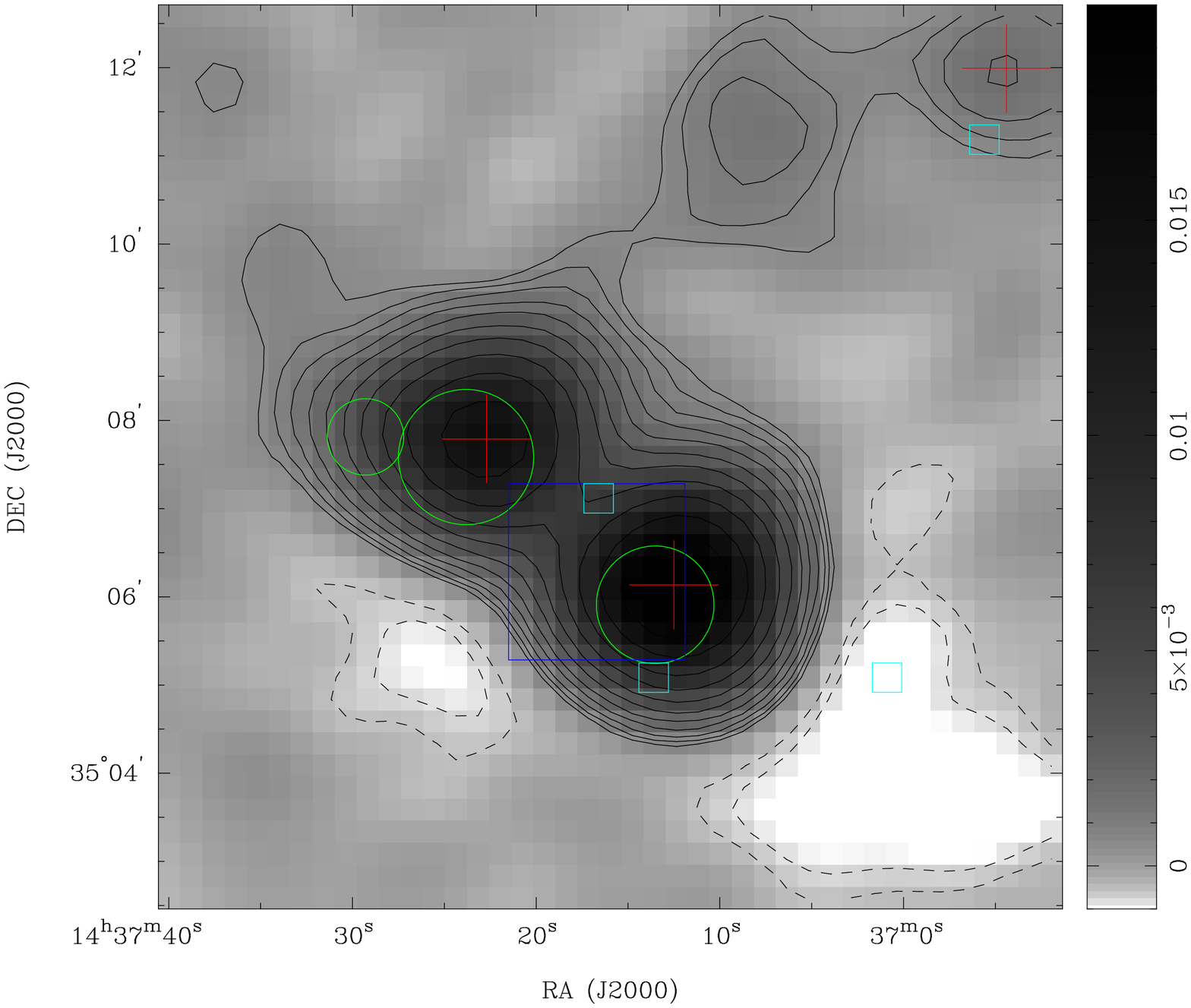,width=0.42\textwidth,angle=-90}}
\mbox{\psfig{figure=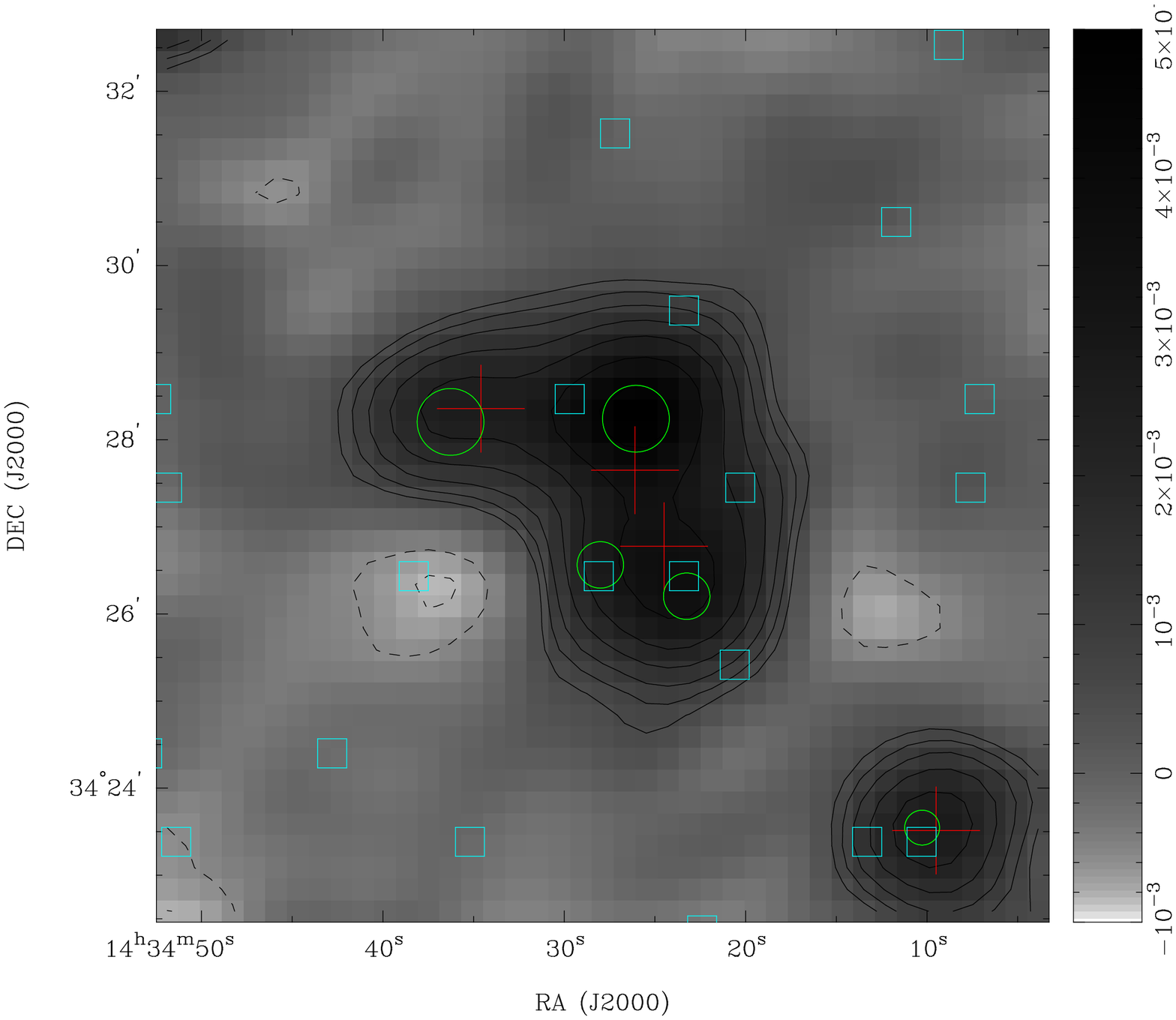,width=0.42\textwidth,angle=-90}\hspace{0.1\textwidth}\psfig{figure=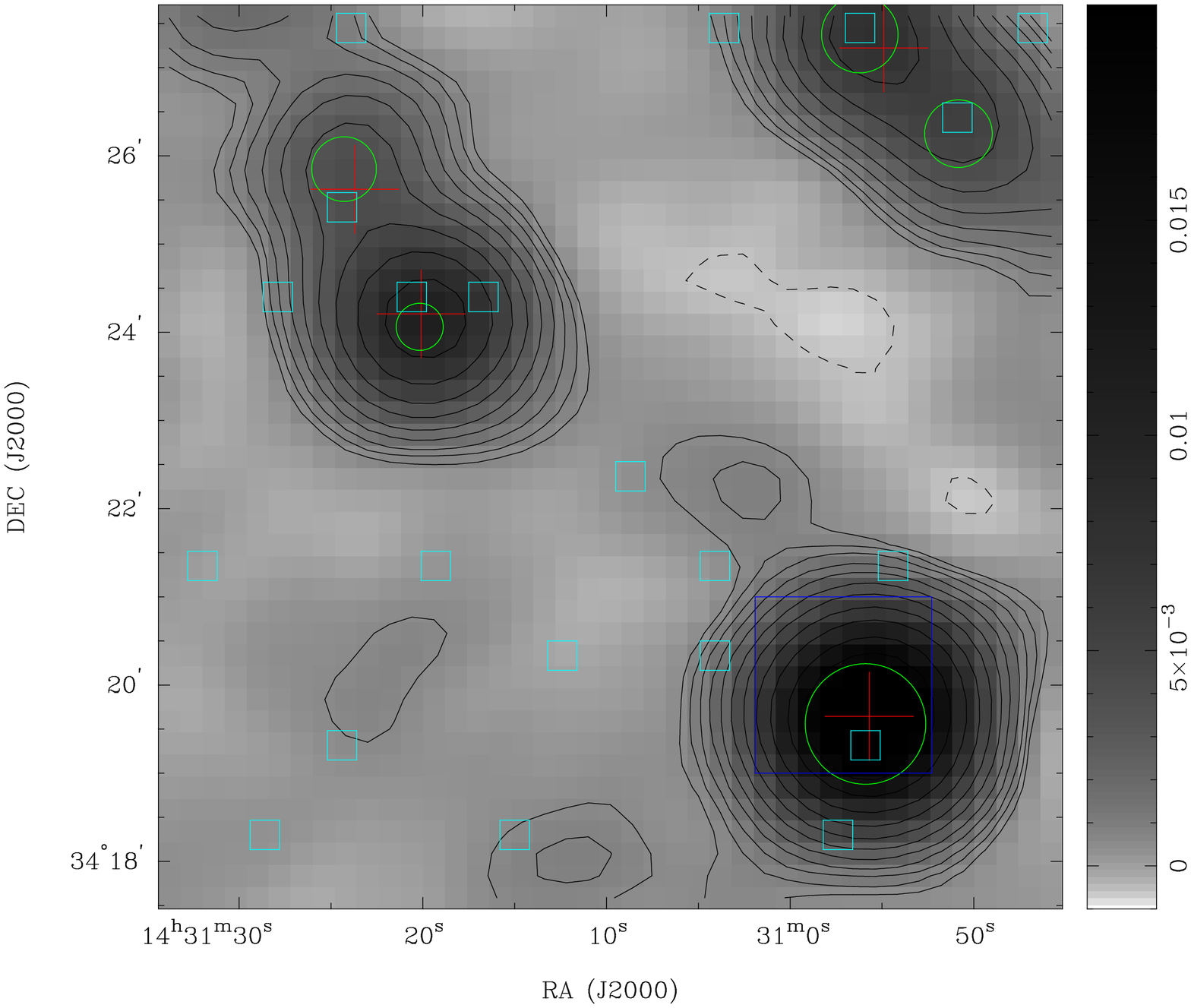,width=0.42\textwidth,angle=-90}}
\mbox{\psfig{figure=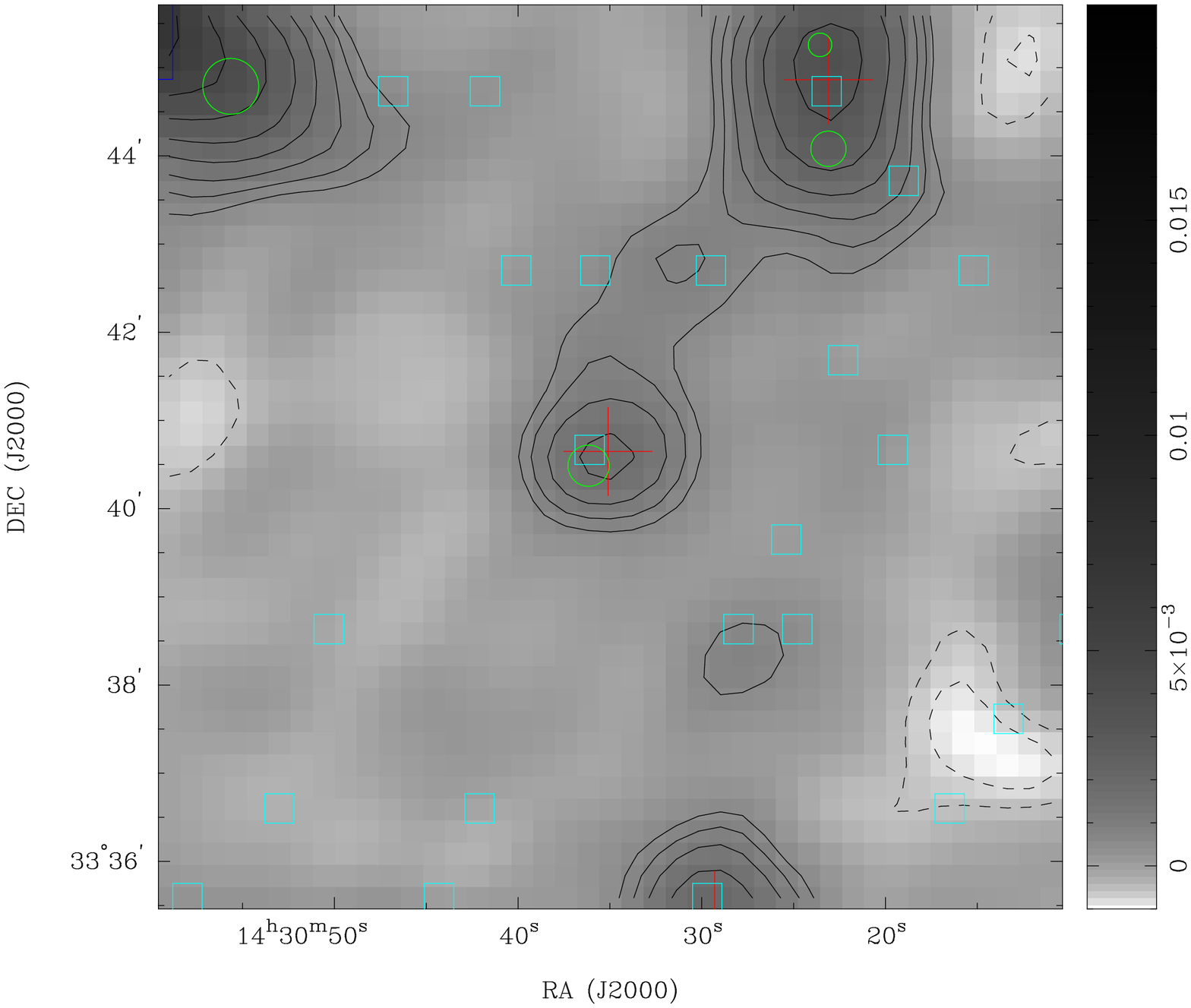,width=0.42\textwidth,angle=-90}\hspace{0.1\textwidth}\psfig{figure=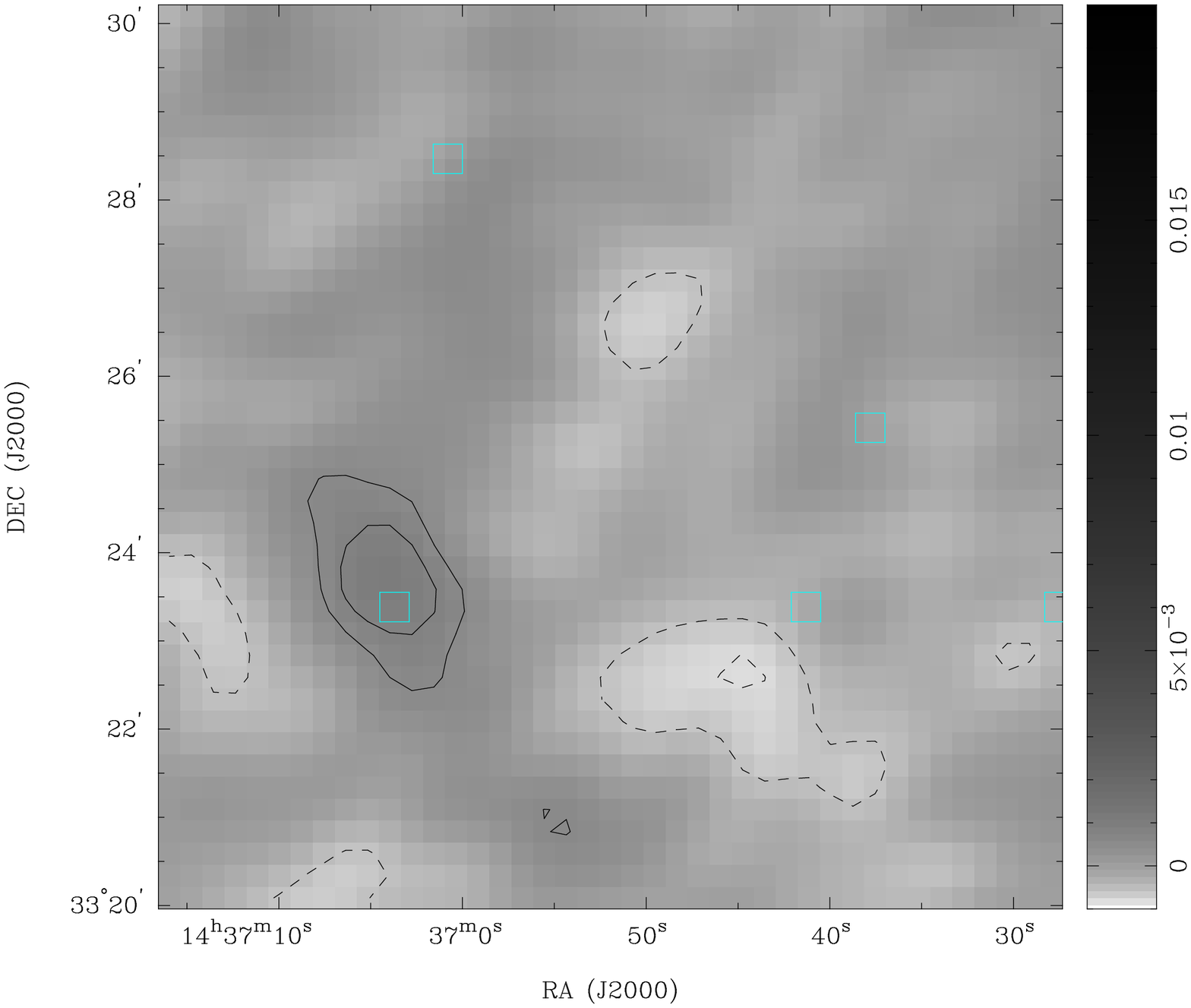,width=0.42\textwidth,angle=-90}}
\caption[]{Images of regions within the large image:  (a) one of the brightest sources in
the field with a flux density of $113 \pm 1$ mJy; (b) a double source; (c) a complex of faint sources; (d) another complex
of sources; (e) the faintest source in the catalog; and (f) an empty field.  All
images have contours that are -4, -2.8, 2.8, 4, 5.6, 8, 11.2, \dots 256 times
0.2 mJy.  The synthesized beam is shown in (a).  Crosses indicate PiGSS sources.
Circles indicate NVSS sources, with area proportional to the logarithm of
the flux density.  Large, blue squares represent GB6 sources.  Small cyan squares
represent WSRT-1400 sources.
Each image is $10\arcmin \times 10\arcmin$.
\label{fig:subim}
}
\end{figure}

We show images of individual sources and small regions in Fig.~\ref{fig:subim}.
Catalog positions for sources from PiGSS and other surveys are indicated.  For sources with
simple structure, we see very good agreement between the PiGSS and NVSS 
catalog decomposition.  The majority of PiGSS sources have this simple
structure.  In the case of more complex source regions such as
Figs.~\ref{fig:subim}c and d, the PiGSS decomposition tends to be simpler and
produces a smaller number of total sources.  The differences are due to the
different resolutions, the more complete $(u,v)$ coverage of the PiGSS data,
the different observing frequencies,
and, possibly, different methods for complex source decomposition.  GB6 sources
are rare and the poorer angular resolution of GB6 leads to even
greater blending of complex structures.

\subsection{Cross-Identifications}

We match PiGSS sources with NVSS sources using a match radius $r_m = 45\arcsec$.  
This radius is chosen to produce an 
an expectation of $N_{false} \lsim 1$ false match between PiGSS and NVSS, assuming a uniform
distribution of sources.  Clustering of sources will
increase the number of false matches.  However, a smaller match
radius may lead to a mismatch between sources that are extended or resolved in 
NVSS.  Approximately 80\% of PiGSS sources have a 
counterpart in NVSS; the remainder are likely to match to sources fainter than
the NVSS flux limit (see the discussion below).
We also match the PiGSS catalog to the GB6 catalog 
using $r_m= 90\arcsec$, leading 
to  $N_{false} \lsim 1$.  Only 8 PiGSS sources have two NVSS sources within
the PiGSS synthesized beam.  Thus, no more than 2\% of NVSS counterparts to PiGSS
sources have their 1.4-GHz fluxes underestimated by source-blending effects.

For the much deeper and higher angular resolution WSRT-1400, 
$r_m=9\arcsec$ will produce $N_{false} \lsim 1$.  As discussed below,
this match radius is comparable to the positional accuracy of our observations;
use of a match radius of this size could lead to a substantial number of PiGSS sources
not being identified.  We select a match radius of $r_m=45\arcsec$ to accomodate
the PiGSS positional uncertainty, leading to $N_{false} \sim 25$.  Note that the
minimum match radius is selected, so the majority of PiGSS matches to WSRT-1400
are closer than this separation.  38 PiGSS sources ($\lsim 10\%$)
have multiple WSRT-1400 matches within the PiGSS synthesized beam. The characteristic
error in the flux density for these multiple-match sources is an underestimate of the
true 1.4-GHz flux density by 16\%, if we assume that multiple-source decomposition does
not create more accurate matches.
The significant differences in source density
and angular resolution between PiGSS and WSRT-1400 indicate that these matches are
best used only when an NVSS or GB6 source is missing.  
We summarize the matches to all surveys
and compute spectral indices ($S \propto \nu^\alpha$)
as a function of frequency for all
of the matches in Table~\ref{tab:matches}.

\begin{figure}[tb]
\psfig{figure=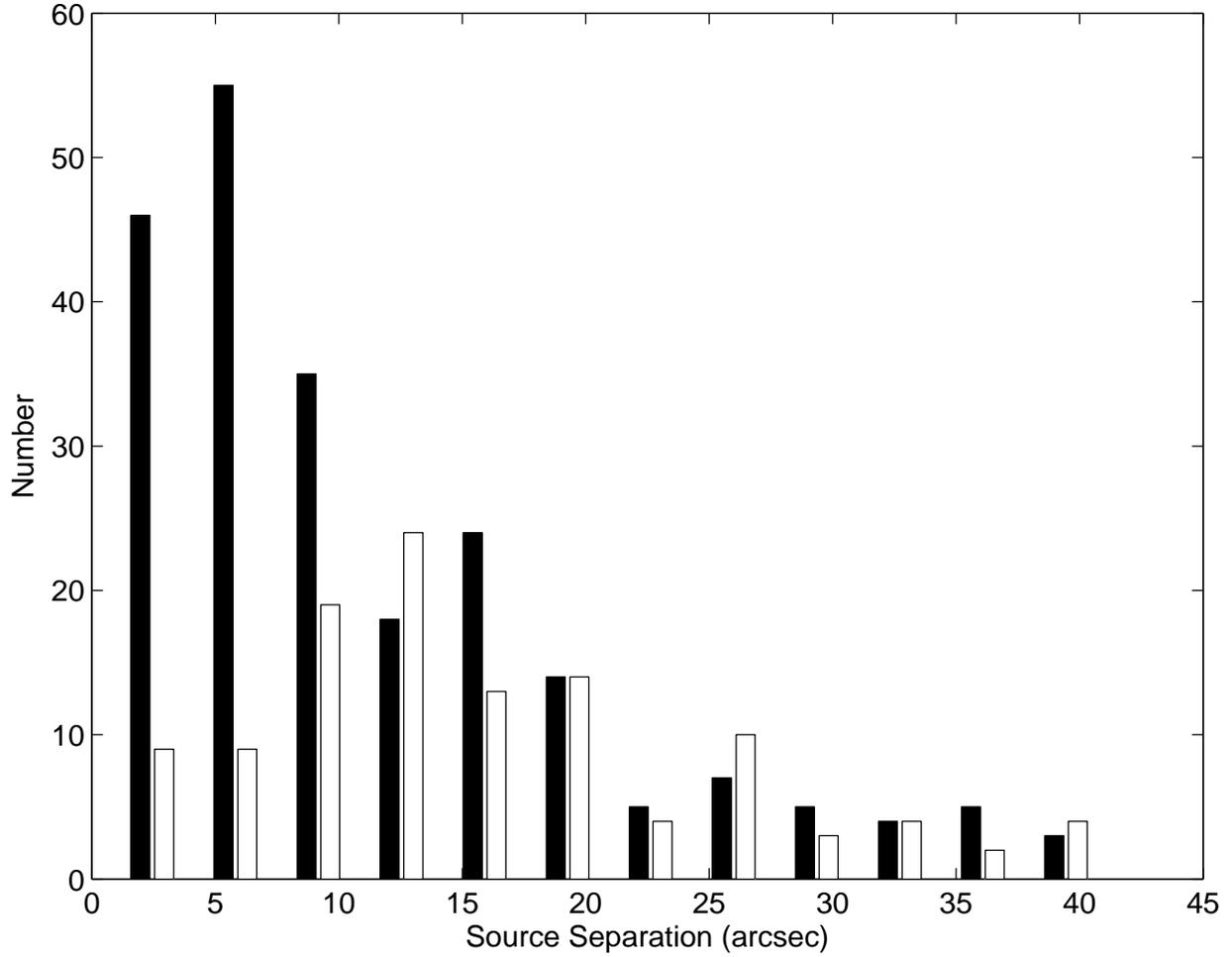,width=\textwidth}
\caption[]{Histogram of separation between PiGSS and NVSS sources.  Solid bars
are for sources within the half-power radii of the outer pointings; open bars 
are for sources beyond that limit.
\label{fig:separation}
}
\end{figure}

\begin{figure}[tb]
\psfig{figure=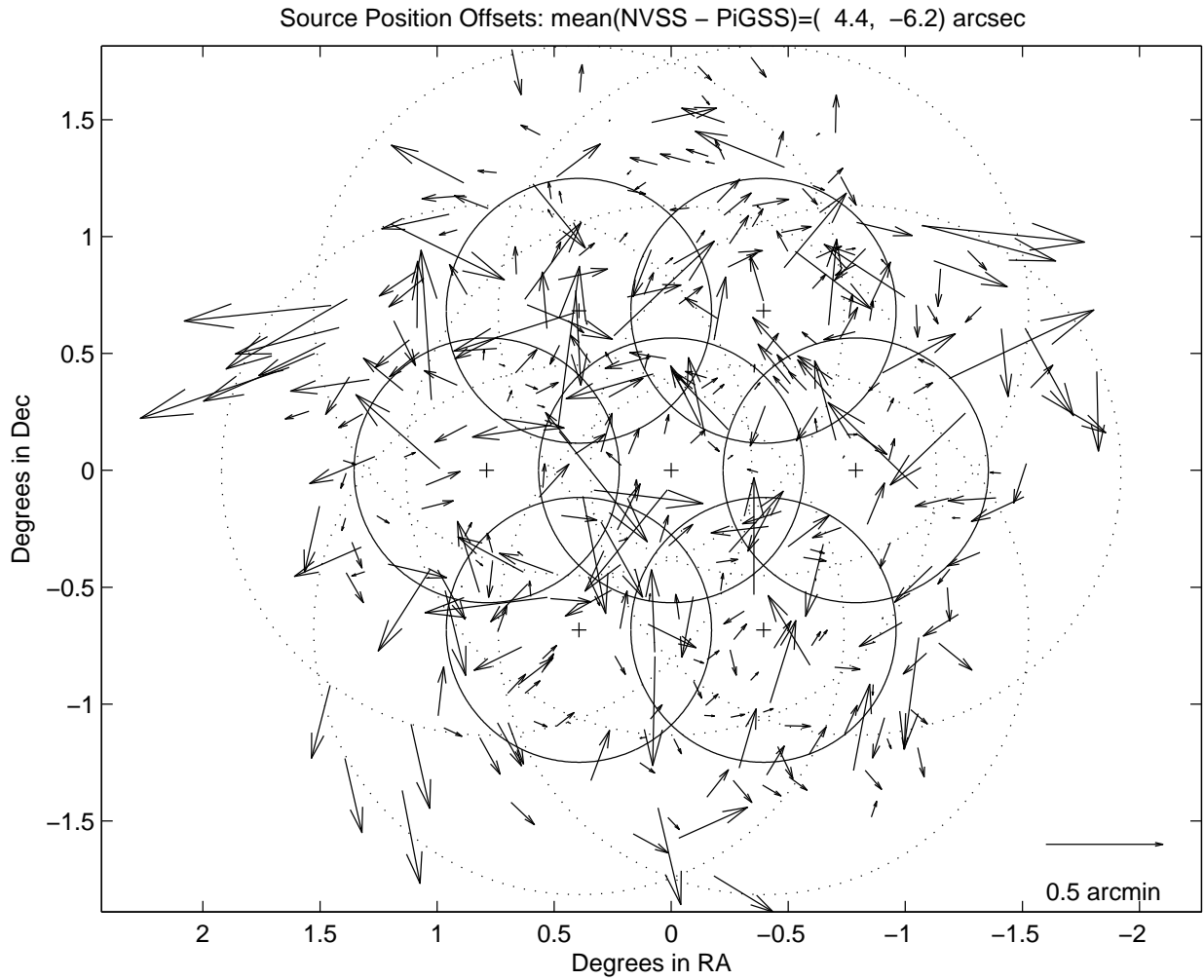,width=\textwidth}
\caption[]{Vectors of position offsets between NVSS and PiGSS.  The mean offset
has been removed.  A scale arrow in
the lower right indicates
0.5\arcmin\ offset.  Note that arrows are much larger than actual scale.
Crosses indicate the centers of each pointing;
solid circles indicate the half-power radii; dashed circles indicate
two times the half-power radii.  
\label{fig:brushedcat}
}
\end{figure}

In Figure~\ref{fig:separation}, we plot a histogram of the separation 
between the PiGSS and NVSS matched sources.  Interior to the half-power radius of the outer pointings,
the median offset in the position is 8.3\arcsec; outside that radius, the
median offset increases to 13.7\arcsec.  There is some evidence for 
systematic shifts in position at radii approaching two times the half-power
radius (Fig.~\ref{fig:brushedcat}).  These shifts are not described as a simple linear shift or a rotation, either around the field center or around individual pointing
centers.
Tests with full 3-dimensional
imaging in CASA appear to indicate that these are not due to phase errors
introduced by ignoring the $w$-term.  Bandwidth and time smearing are not 
significant at these radii.  
These localized shifts may be due to primary-beam phase errors at radii beyond the
half-power point.  
The accuracy of matching with WSRT-1400 sources
is comparable to that achieved with NVSS; the median separation between PiGSS and WSRT-1400
sources in 10.5\arcsec.

\begin{figure}[tb]
\psfig{figure=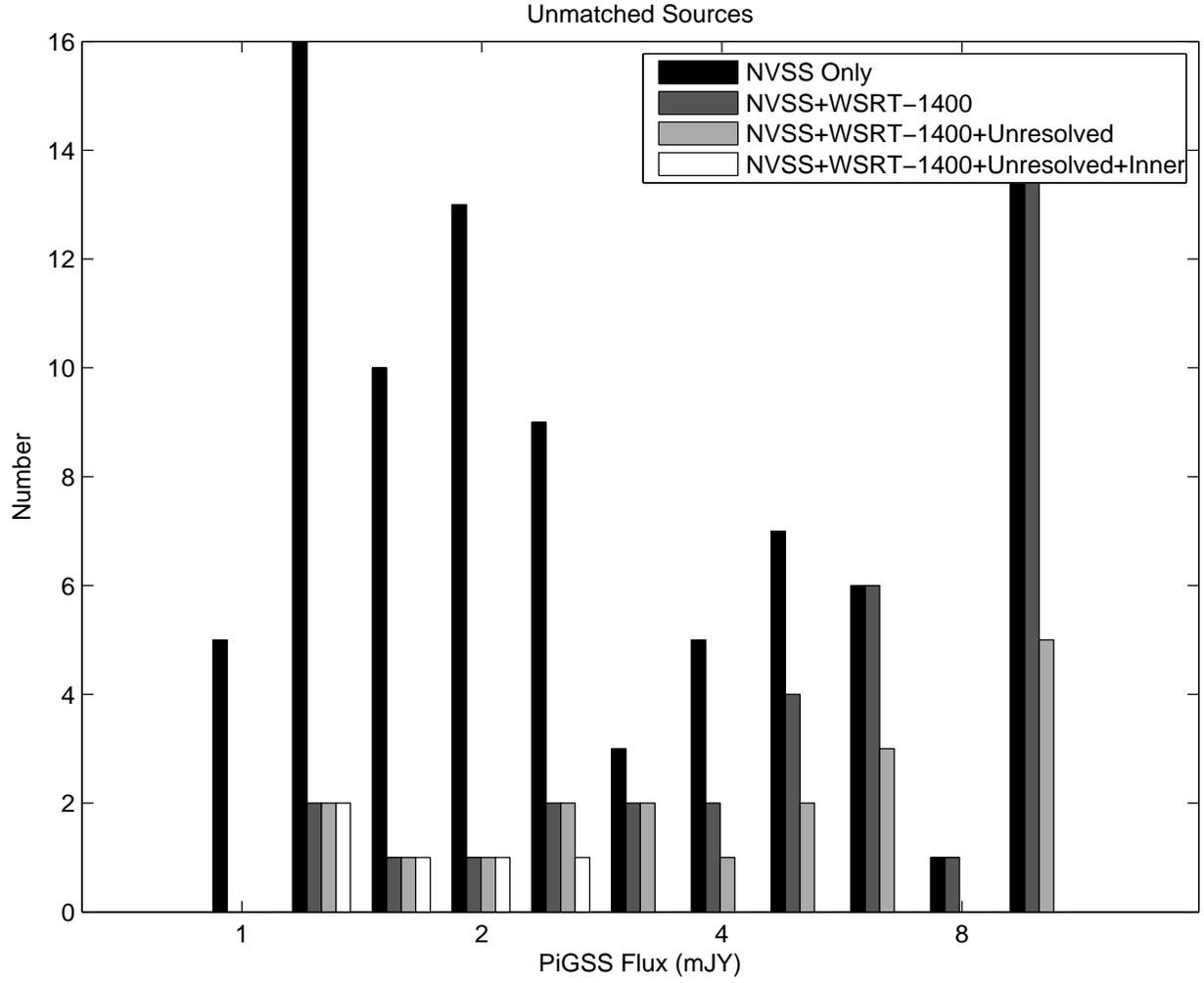,width=\textwidth}
\caption[]{Histogram of PiGSS sources not identified in matches with other 
catalogs.  See text for details.
\label{fig:uid}
}
\end{figure}

\begin{figure}[tb]
\psfig{figure=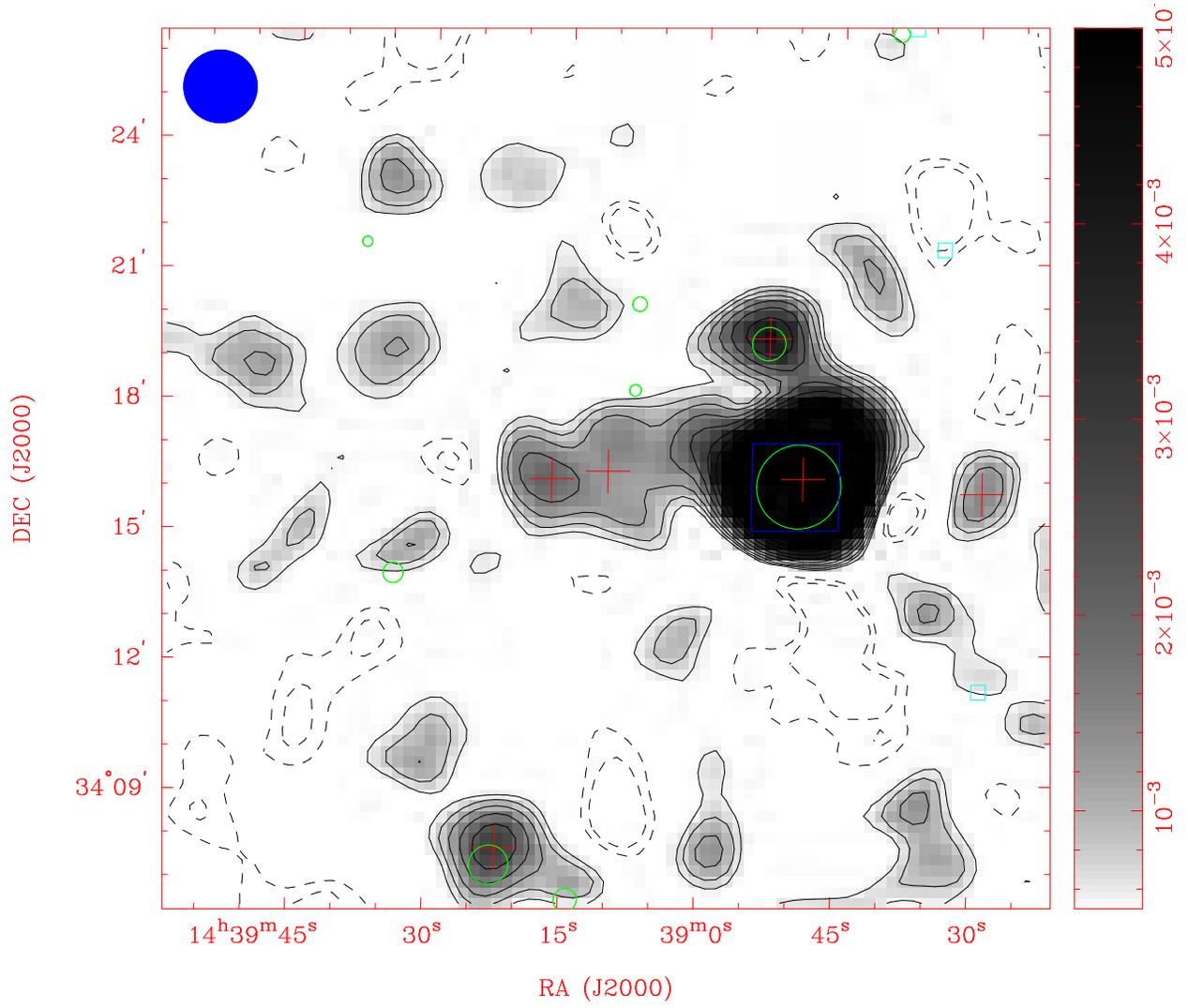,width=\textwidth,angle=-90}
\caption[]{An extended PiGSS source not observed in NVSS.
Contours are the same as in Fig.~\ref{fig:subim}.
\label{fig:extended}
}
\end{figure}

\begin{figure}[tb]
\mbox{\psfig{figure=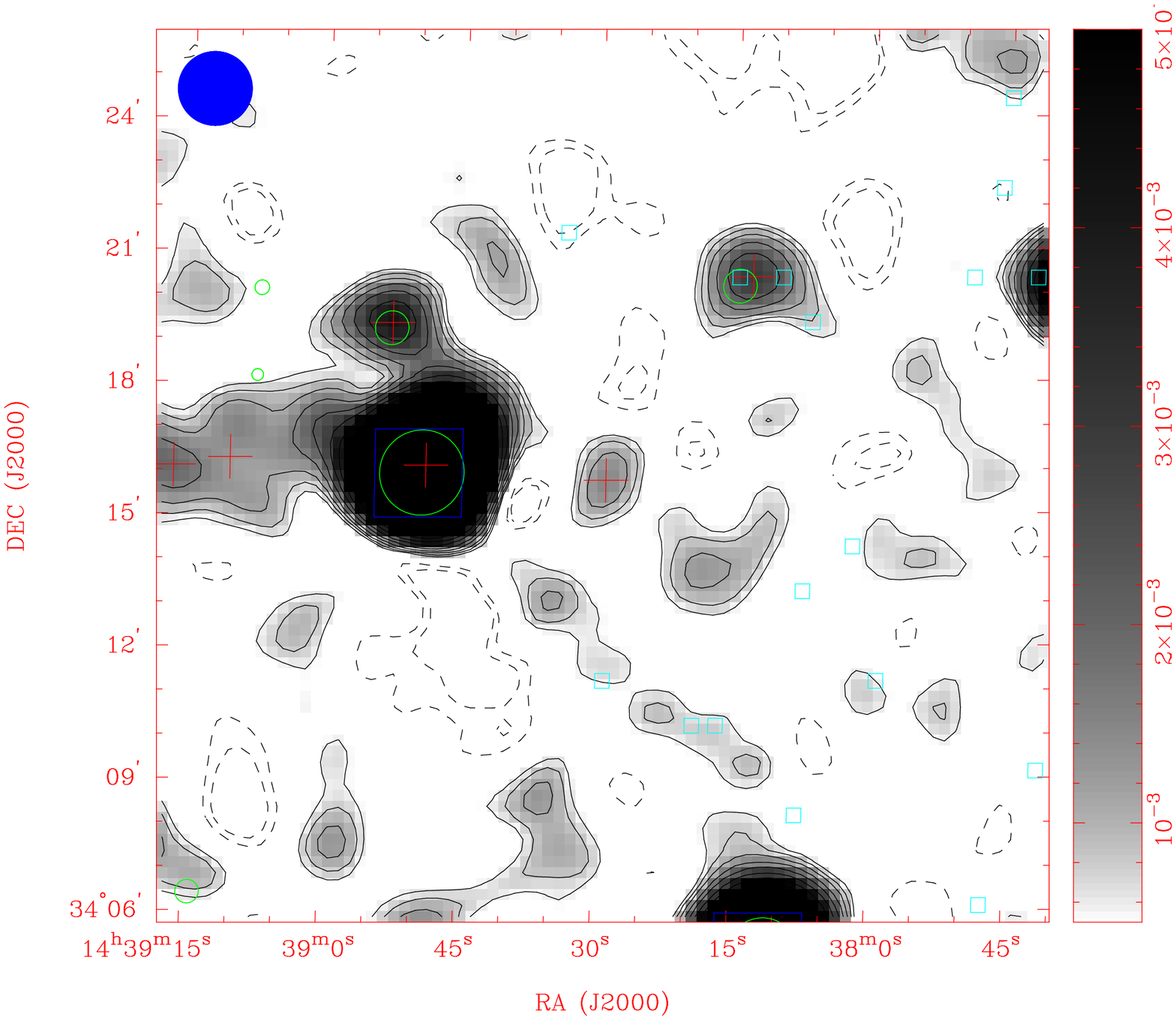,width=0.42\textwidth,angle=-90}\hspace*{0.1\textwidth}\psfig{figure=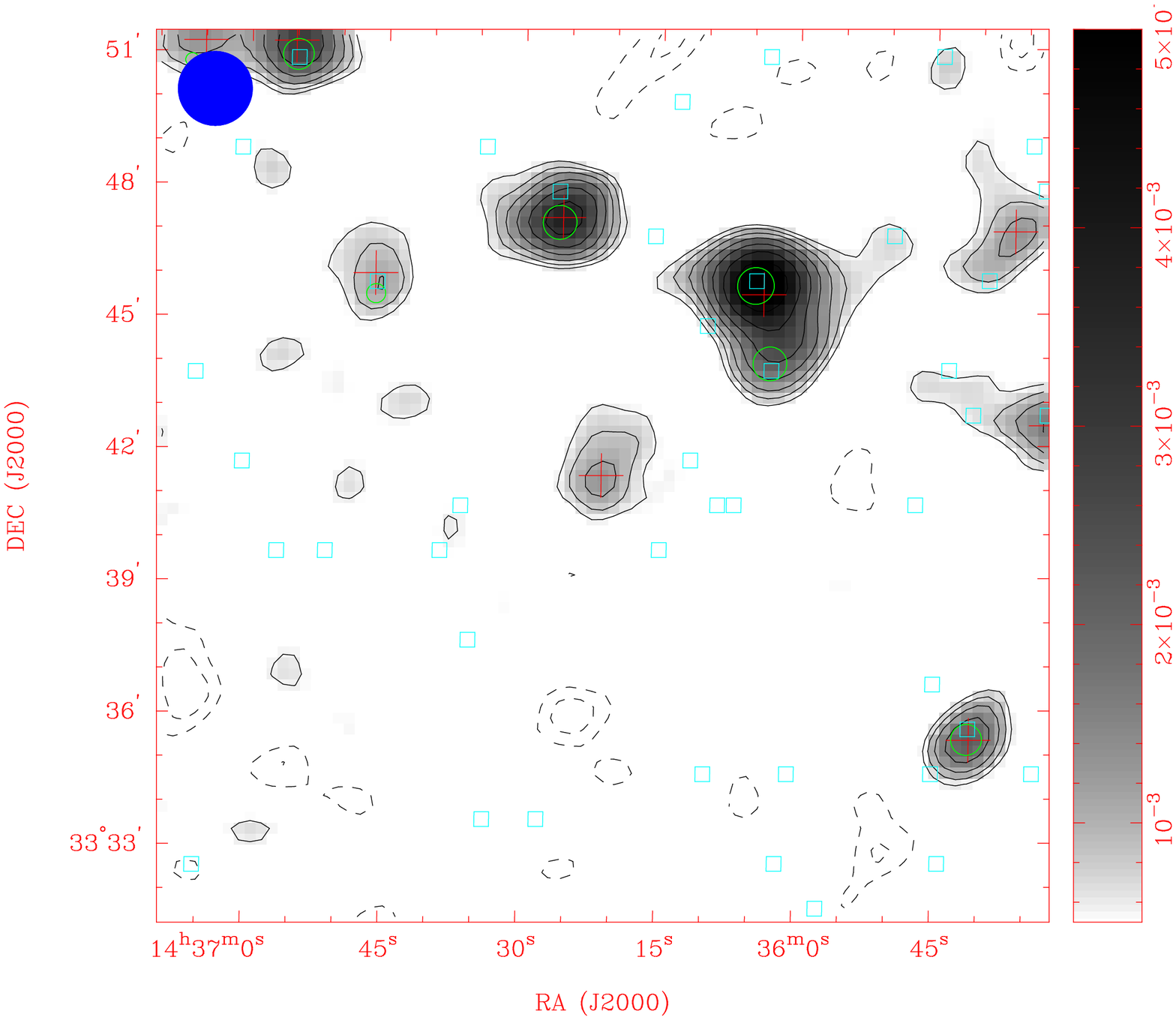,width=0.42\textwidth,angle=-90}}
\mbox{\psfig{figure=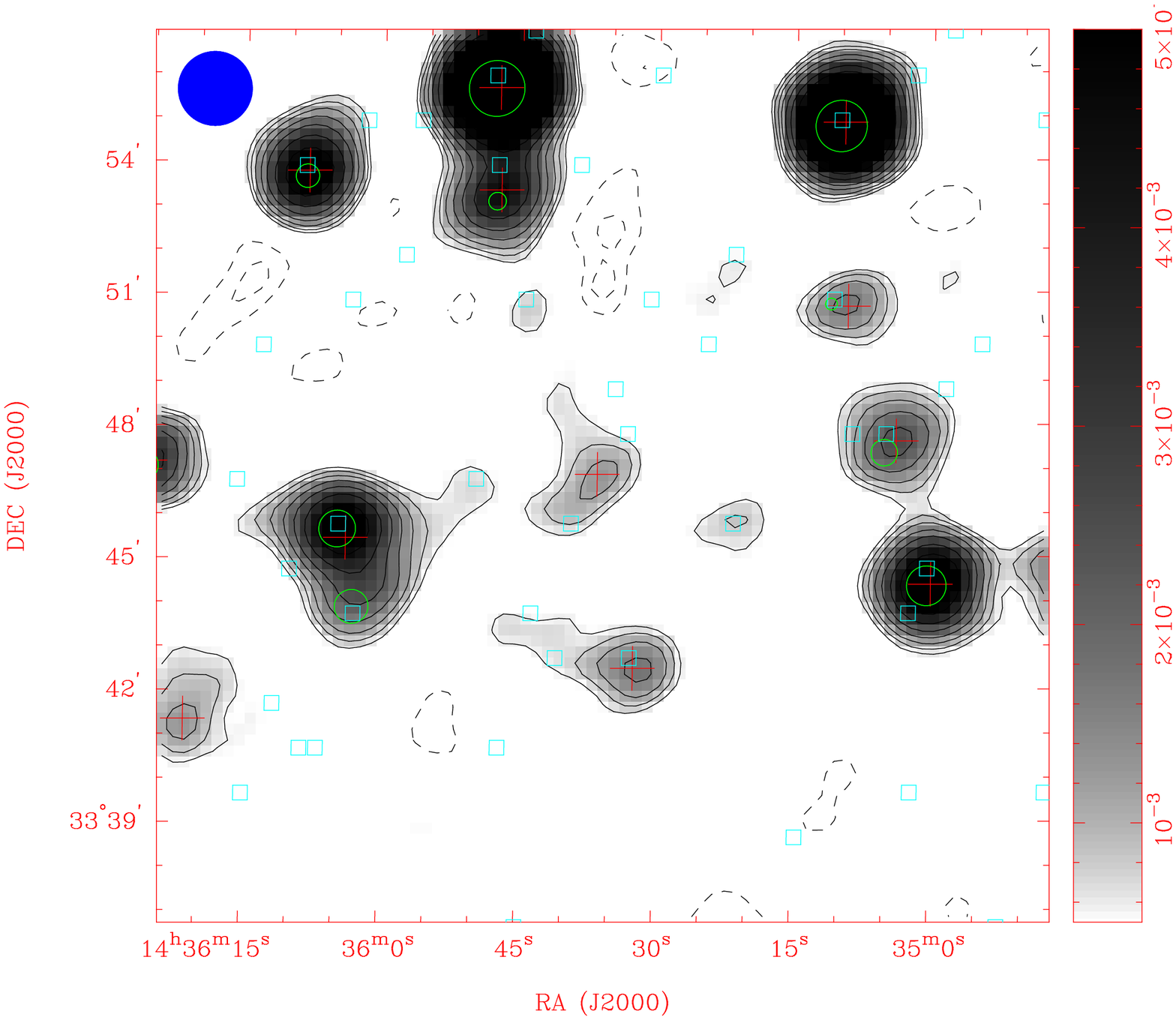,width=0.42\textwidth,angle=-90}\hspace*{0.1\textwidth}\psfig{figure=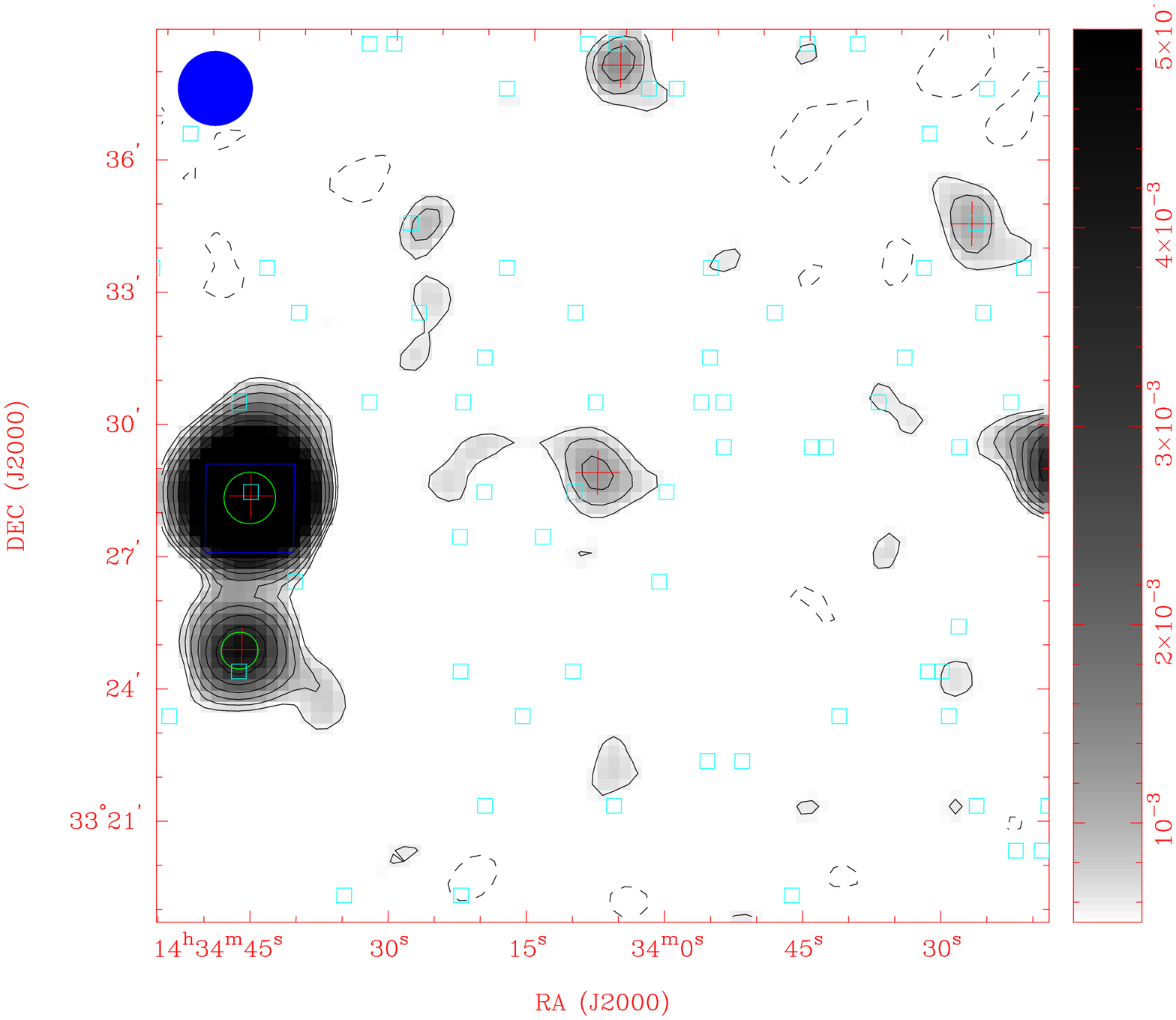,width=0.42\textwidth,angle=-90}}
\mbox{\psfig{figure=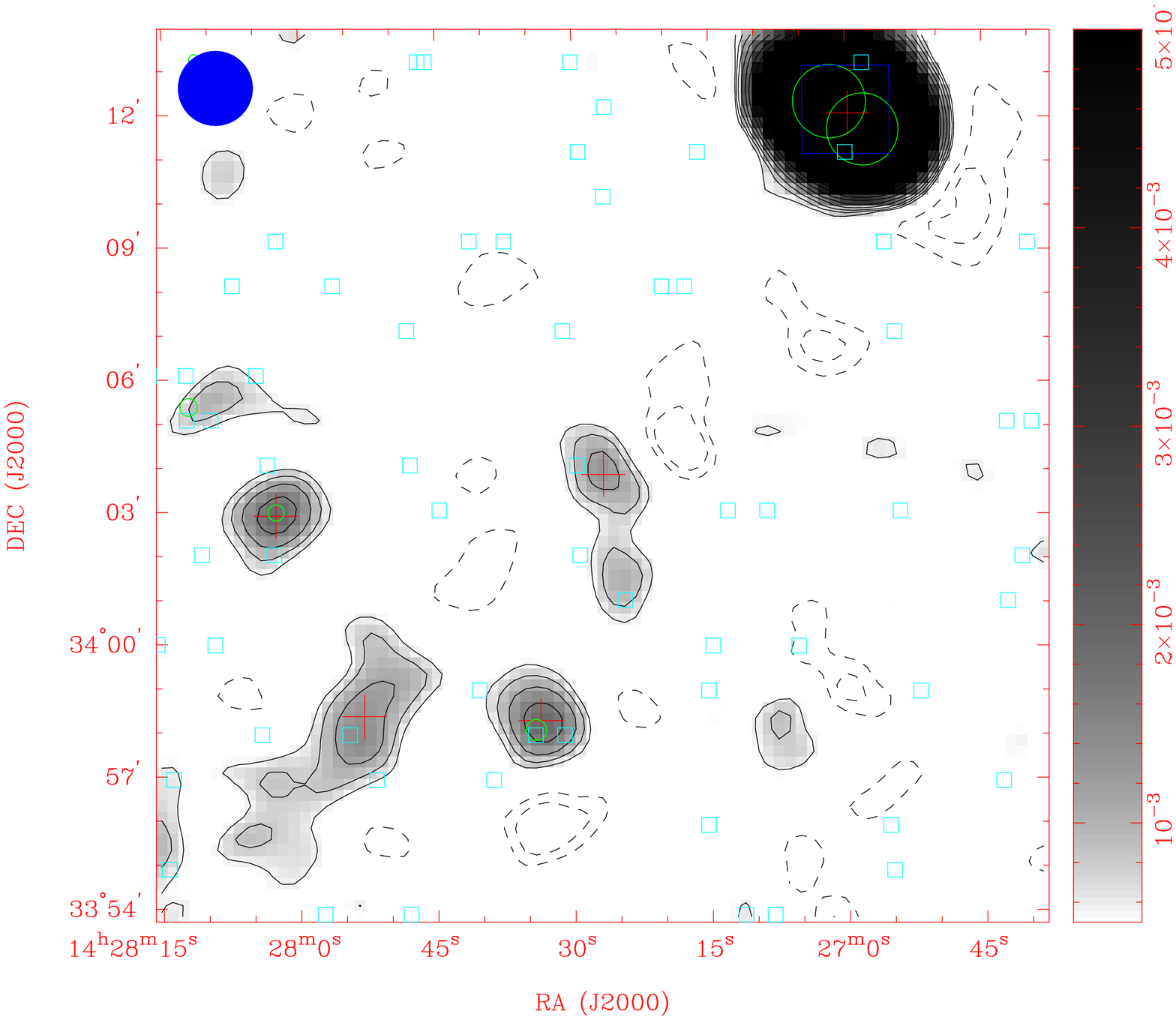,width=0.42\textwidth,angle=-90}}
\caption[]{Images of unresolved PiGSS sources not identified in NVSS and WSRT-1400 and
within the half-power contour of the image.  The source without an identification is 
located at the center of each image.
\label{fig:nomatch}
}
\end{figure}

We use multiple catalogs to identify PiGSS sources without associations  (Fig.~\ref{fig:uid}).  Absence of
an NVSS counterpart is unsurprising for low flux density sources given the lower 
detection threshold for PiGSS; 89 PiGSS sources are unmatched to NVSS.  
If we match against the WSRT-1400 survey as well as
NVSS, we are able to identify more sources, leaving us with 35 unidentified
sources.  Many
of the remaining unmatched PiGSS sources are outside the boundary of the WSRT-1400
survey; i.e., these sources could only be matched to NVSS.  The different
resolutions and $(u,v)$ coverages of these surveys also leads to mismatches.
For instance, we see extended structure associated with a bright source that 
does not appear in the NVSS image of the field (Fig.~\ref{fig:extended}).  This resolved
source is represented by the spike at $>8$ mJy  in the histogram.  If we further exclude all
resolved sources from the histogram, we are left with sources with
flux densities less than 2.5 mJy.  Further excluding sources outside the half power
sensitivity, we find five sources with flux densities of $\sim 1.5$ mJy that 
have no match in any of the catalogs.  We show images of these sources in 
Fig.~\ref{fig:nomatch}.  Of these sources, one is adjacent to a very bright
source with resolved structure and is possibly a sidelobe of that source or is missed in other surveys due to confusion with that bright source; three have 
probable WSRT-1400 matches that are beyond the 45\arcsec\ match radius.  Only 
J143621+334120 has no apparent match and no reason to exclude it as a real source.

All NVSS sources with flux densities greater than 5 mJy are matched to PiGSS sources
either through a clear one-to-one match or with an association with a resolved 
source that has a different decomposition into components in PiGSS and NVSS.
53 NVSS sources with flux densities between 2.5 and 5 mJy
are not matched to PiGSS sources; this is primarily due to
a combination of faint 3.1 GHz emission that is not detected in PiGSS, source
confusion, and a few sources that are missed by the PiGSS detection algorithm.

All but two GB6 sources have matches to PiGSS sources.  GB6 source J143921+344803
falls in between two PiGSS sources with a separation of $\sim 4$\arcmin\ and is
probably a blend of these two.  GB6 source J142554+343552, which has
a flux density of 21 mJy, has no 
apparent counterpart in PiGSS, NVSS, or WSRT-1400.  
The nearest PiGSS source is $\sim 5$\arcmin\ away.
The field contains a number of nearby galaxies identified by SDSS but there are no
AGN or known compact objects.
There is no known reason to exclude this GB6 source as a transient source but 
without deeper access to the GB6 data we cannot determine whether it is real
or spurious.

\subsection{Spectral Indices}

\begin{figure}[tb]
\psfig{figure=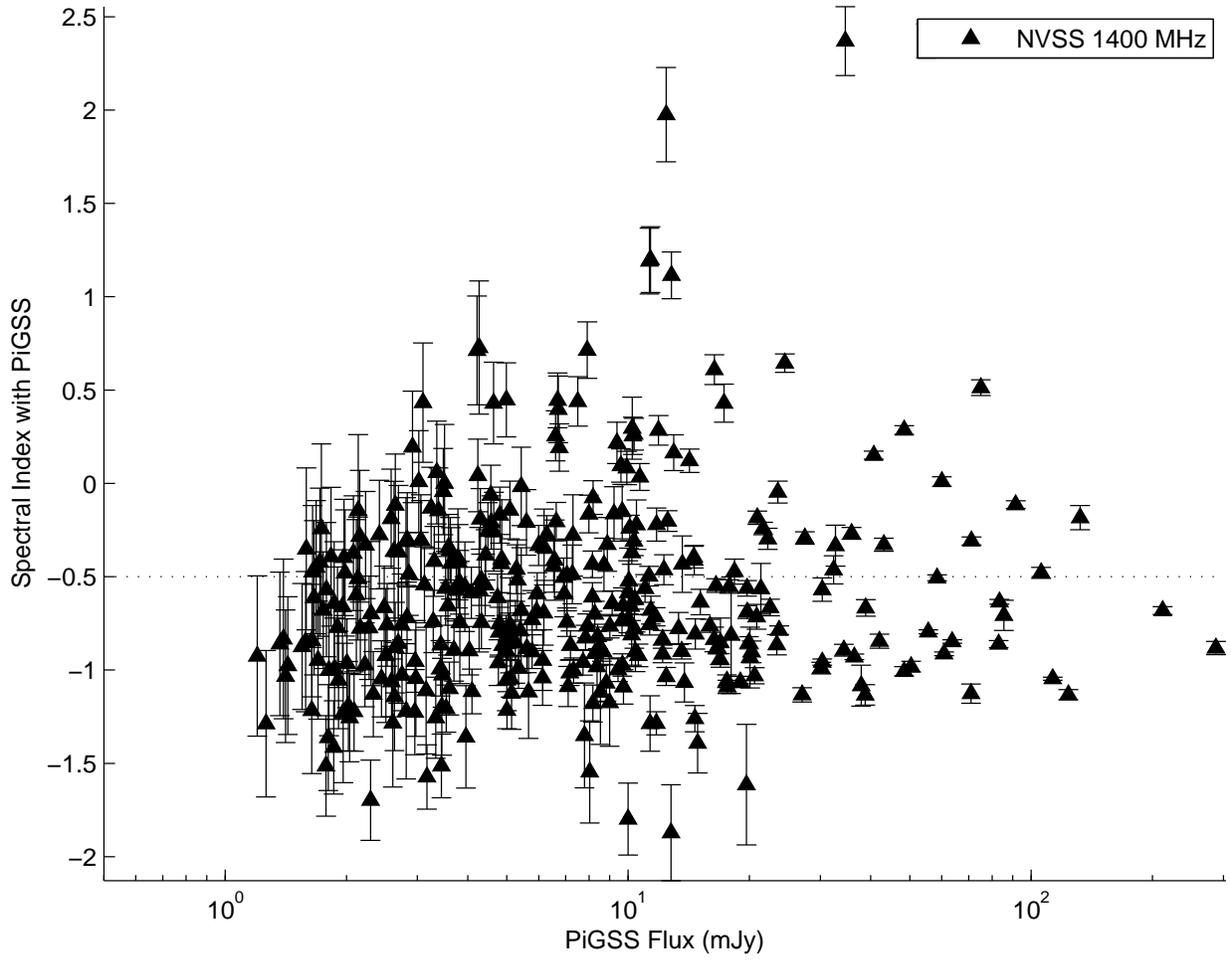,width=\textwidth}
\caption[]{Spectral indices generated from PiGSS 
with respect to the NVSS.
The dashed line indicates the cutoff of $\alpha > -0.5$ for flat spectrum
sources.
\label{fig:specindex}
}
\end{figure}

Spectral indices are tabulated in Table~\ref{tab:matches} and
plotted in Fig.~\ref{fig:specindex}.  There is good agreement between 
spectral indices determined with NVSS and WSRT-1400 data.  The median 
spectral indices with the two surveys are -0.68 and -0.65, respectively.  
Excluding 25 sources that have very large spectral indices in one or
the other survey (and are, therefore, likely to be mismatched), the
rms difference in spectral index relative to NVSS and WSRT-1400 is 0.17.
This is an upper limit on the systematic error in the NVSS spectral indices 
because it includes relative calibration error between NVSS and WSRT-1400.
The drop in median spectral index at low flux densities is a selection
effect due to the 2.5-mJy flux limit of the NVSS catalog; 
inverted or flat-spectrum PiGSS sources will not be detected in NVSS at low flux density.
GB6 spectral indices have a larger scatter,
caused by larger error in the GB6 flux densities, the coarser
angular resolution of GB6, and, possibly, higher variability among these sources.  
Only one GB6
source (out of 37) has a PiGSS source within 90 arcsec radius, which is well-beyond the
half-power radius of ~70 arcsec.  Thus, we are not significantly affected by source-blending effects
in the PiGSS-GB6 spectral indices.

\subsection{Number Counts}

\begin{figure}[tb]
\psfig{figure=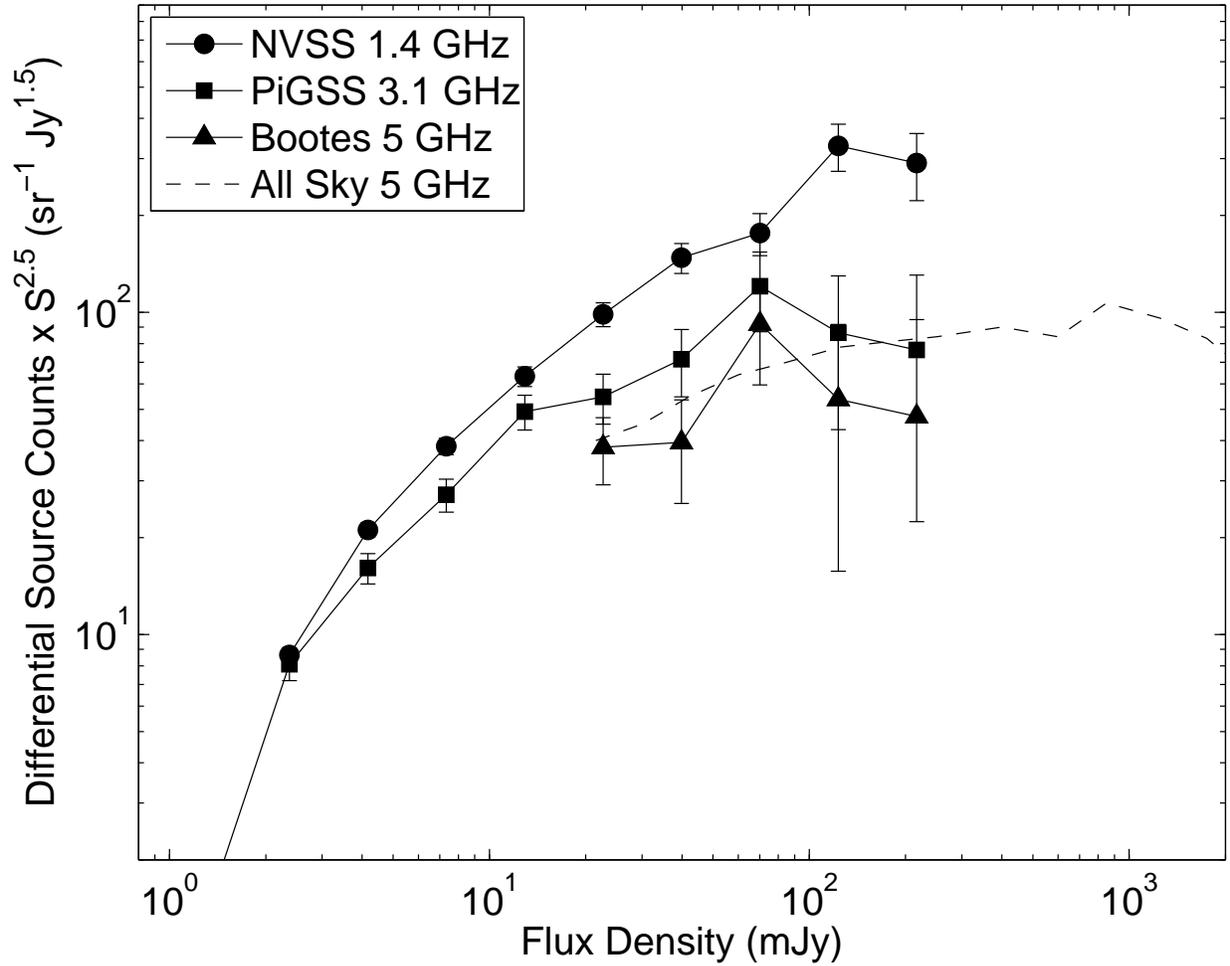,width=\textwidth}
\caption[]{Number counts as a function of flux density from PiGSS data (squares),
NVSS sources in the same field (filled circles), GB6 sources in the same field
(filled triangles), and 5 GHz source counts from the full GB6
catalog (dashed line; Gregory et al. 1996).
\label{fig:numcounts}
}
\end{figure}

We construct number counts for the PiGSS, NVSS, and GB6 sources in the field 
(Fig.~\ref{fig:numcounts}).  As expected, PiGSS number counts fall in between
the NVSS and GB6 number counts
since the average spectral index is $\sim -0.7$.  This result indicates that we are
detecting the expected population in this field.

\section{Discussion}

\subsection{Variable and Transient Sources}

A major goal of PiGSS is to identify and characterize variable and transient radio
sources.  The results of this paper, specifically the catalog of sources in
comparison with existing radio catalogs of the richly surveyed Bo\"{o}tes field, enable us
to investigate variability with a characteristic time scale of $\sim 4$ months
relative to earlier snapshot epochs.  That is, transient
sources that appear in the PiGSS data are most likely to have a lifetime comparable to
the total integration time of this survey; variable sources can have evolved even
more slowly, i.e., over the course of years from prior epochs to the present.
Brighter, shorter-duration transients are possible but less probable if they 
have source populations that increase with decreasing flux density.
The reverse analysis --- identification of sources present in NVSS, WSRT-1400,
or other survey but not present in PiGSS --- provides an opportunity to probe source
populations with characteristic timescales that match the
shorter snapshot times of those surveys.  In the case of NVSS, that time scale is
minutes.  

As discussed above, we identify one source J143621+334120 that appears in PiGSS
but in none of the other catalogs.  With a flux density of $1.80 \pm 0.42$ mJy, the
source is a marginal detection at 4.3$\sigma$.  For purely Gaussian image noise,
the expectation over the entire image is of $\sim 0.1$ sources at this threshold or above.
Thus, it's not unlikely that this is a random fluctuation or the
result of increased noise threshold from systematic error.   In a forthcoming paper we 
will investigate the variability of this source in daily and monthly images.

\begin{figure}[tb]
\psfig{figure=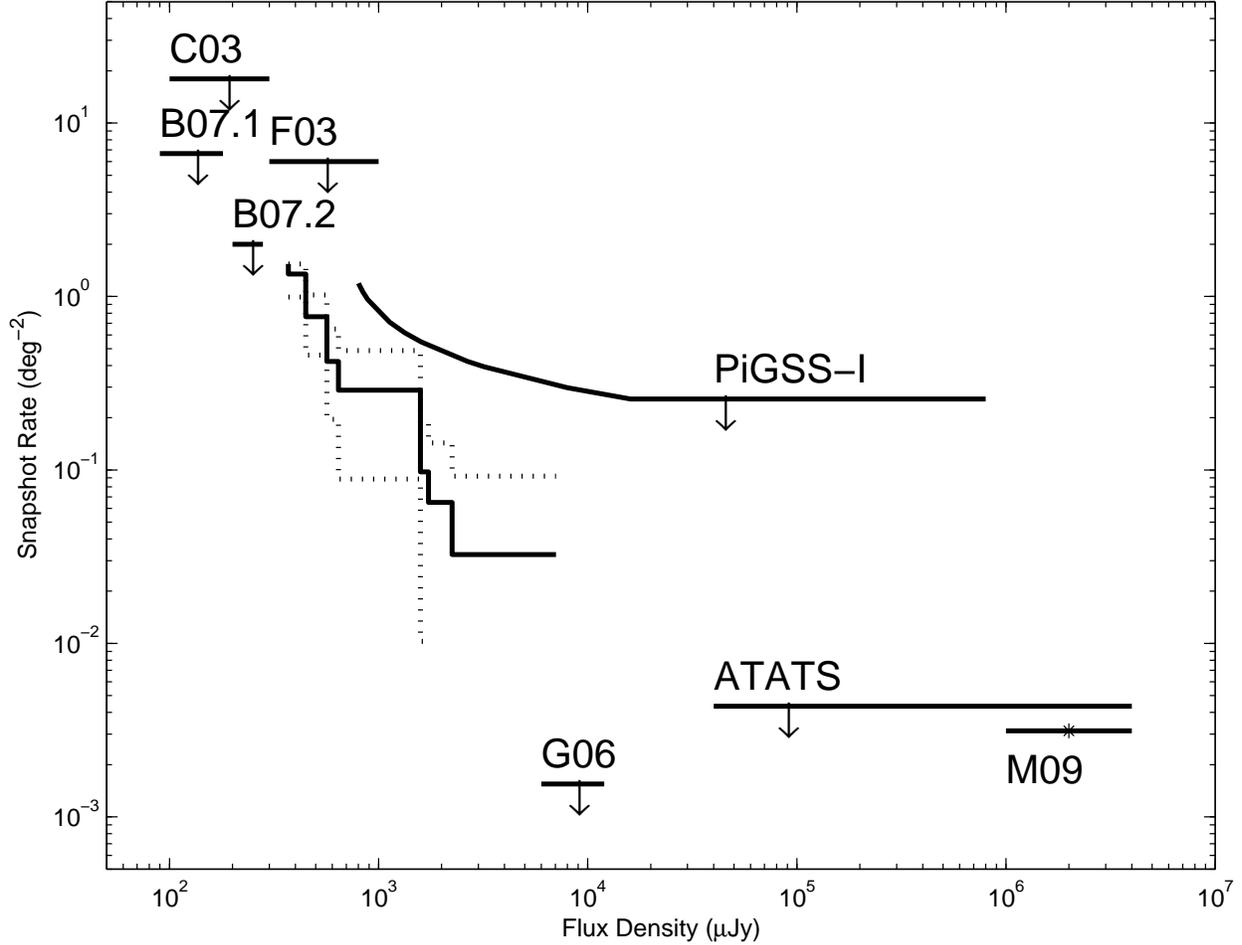,width=\textwidth}
\caption[]{Two-epoch transient rates from PiGSS and other surveys as a function
of flux density.  The curved solid black line labeled PiGSS-I
is the limit from this paper.
The solid black line with step functions shows the rate from 
\citet{2007ApJ...666..346B} while the
dotted lines show the $2\sigma$ upper and lower bounds. The arrows show
$2\sigma$ upper limits for transients from 
\citet{2007ApJ...666..346B} with a
1-year timescale (B07.1), two-month timescale (B07.2), and for transients
from the comparison of the 1.4~GHz NVSS and FIRST surveys
\citep[G06;][]{2006ApJ...639..331G}, from the \citet{2003ApJ...590..192C}
survey (labeled C2003), from the \citet{2003AJ....125.2299F}
survey (labeled F2003), from ATATS \citep{2010ApJ...719...45C},
and from the \citet{2009AJ....138..787M} survey (labeled M09).  
\label{fig:rates}
}
\end{figure}

Since we do not have firm detection of any transient sources, we do not have
a transient timescale.  In this case, the snapshot rate $R$ (Eqn.~\ref{eqn:snapshot}), which 
compares the number
of transients in a two epoch survey is the appropriate statistic to make use of.
We, therefore, set an upper
limit to the transient rate for long-duration transients, $R_{month}\lsim 1$ deg$^{-2}$
at 1 mJy and $\lsim 0.3$ deg$^{-2}$ at 10 mJy.
In the reverse case of identifying NVSS sources
missing in PiGSS, we set a limit $R_{minute} \lsim 0.3$ deg$^{-2}$ for flux densities
greater than 5 mJy.  Exact limits for long-duration transients are plotted in 
Fig.~\ref{fig:rates} 
along with rates from past observations.

Rates on transient and variable sources will  improve significantly with the release of 
daily monitoring results for the Bo\"{o}tes field and yearly monitoring for the NGP PiGSS catalogs.
For the former, we anticipate two orders of magnitude greater sensitivity to 
short-timescale 
phenomena at flux densities an order of magnitude higher.  For the latter, 
we anticipate three orders of magnitude improvement in transient and variable rates.

The current set of limits for transients of duration 4 months place the most interesting
limits on AGN, ESEs, and RSNe, all 
of which can have durations of months to years.  The case of III Zw 2 is instructive 
for extreme AGN variability; it exhibited a factor of $>20$ outburst rising from below
100 mJy to greater than 1 Jy on a timescale of
1 year \citep{1999ApJ...514L..17F,2005A&A...435..497B}.  III Zw 2 
has exhibited episodic variability of this kind with a duty cycle of 
$\sim$ 20\%. 
Such a source would be readily detectable in this survey down to our detection threshold of 1 mJy.
Thus, we can limit the number of III Zw 2 analog sources to be $\lsim 0.2$\% of all radio
sources.  ESEs have timescales of months and amplitudes of order factor of 2 and therefore are harder
to distinguish in a two-epoch comparison.  Detailed analysis of daily light curves will be sensitive
to ESE behaviour.

\subsection{Flat Spectrum Sources}

Of particular interest is the construction of a flat spectrum catalog from
the PiGSS data.  We identify 124 sources with $\alpha > -0.5$
and 39 sources with $\alpha > 0$.  
If we restrict our sample to those within the half-power radius where fluxes are 
more accurate, then we find 26 sources with $\alpha > 0$, or 8 per square degree.

A single source from the LAT 1-year catalog falls in the Bo\"{o}tes field, 1FGL J1426.0+3403
\citep{2010ApJS..188..405A}.  This
source is a probable match to the PiGSS source J142607+340433, which has a flux of 
$27.5 \pm 0.6$ mJy.  Matches to the other catalogs indicate a spectral index of $\sim -0.2$.
The source is not identified in 
the CGRaBS catalog of 
flat spectrum radio sources, presumably because it falls below the detection threshold 
for that catalog
\citep{2008ApJS..175...97H}.  
The LAT catalog associates this source with the BL Lac object, BZB J1426+3404
\citep{2009A&A...495..691M}.
As the LAT catalog becomes more sensitive and detects more transient and variable sources, we
can expect an increased number of LAT sources to fall into these fields.
The next PiGSS paper will study short-term variability of sources in the field,
which may be a stronger predictor of gamma-ray activity than spectral index.

The BZ catalog identifies 3 blazars, BL Lac objects, 
and radio loud quasars in the Bo\"{o}tes field, using 1.4 GHz radio flux
data and optical spectra
\citep{2009A&A...495..691M}.
The BZ catalog has an average density of $\sim 0.1$
sources per square degree.  The higher number density in the Bo\"{o}tes
field is due to the non-uniformity of the available optical spectroscopy.
The $\sim 10^2$ flat spectrum sources identified 
via PiGSS spectral indices indicates an order of magnitude increase
in the number of blazar candidates.  Over the $10^4$ square degrees of
PiGSS, we expect to identify $\sim 10^4$ flat spectrum sources above a
flux density of 10 mJy.  This is an order of magnitude increase over the
1625 flat spectrum sources identified in CGRaBS.
Comparison of PiGSS sources  in Bo\"{o}tes with
the extensive NDWFS imaging and spectroscopic results should provide a 
compelling characterization of the flat spectrum population in this field.

\section{Conclusions}

We present the first results from the Pi GHz Sky Survey in this paper.  This includes a
description of the overall survey strategy and goals with an emphasis on transient
and variable source characterization.  The data presented in this paper represent 
less than 3\% of the total data that will be obtained in PiGSS.  These data
are a deep image of a region that is approximately 10 deg$^2$ in the Bo\"{o}tes
field.  Comparison of the PiGSS results with existing surveys of the field, especially
NVSS, provides an important verification of the fidelity of the PiGSS data.  
We detect $\sim 10^2$ flat spectrum sources in this early stage of PiGSS.
We identify one PiGSS source that does not appear in any other radio catalog.  This
source is possibly transient or strongly variable but may also be due to 
fluctuations in the noise; a future paper will examine the daily and
monthly images of this source.
We set upper limits on transients with lifetimes of
minutes and months.  The complete PiGSS catalog will provide a significant 
exploration of transient parameter space on time scales that range from minutes to years.

\acknowledgments

The authors would like to acknowledge the generous support of the Paul
 G. Allen Family
 Foundation, who have provided major support for design, construction,
 and operations of
 the ATA. Contributions from Nathan Myhrvold, Xilinx Corporation, Sun
 Microsystems,
 and other private donors have been instrumental in supporting the ATA.
 The ATA has been
 supported by contributions from the US Naval Observatory in addition
 to National Science
 Foundation grants AST-050690, AST-0838268, and AST-0909245.
We dedicate this paper to the memory of the late Don Backer: mentor, colleague,
and friend.

\bibliographystyle{apj}
\bibliography{myrefs}

\begin{deluxetable}{lllll}
\tabletypesize{\scriptsize}
\tablecaption{PiGSS Fields and Specifications \label{tab:pigss}}
\startdata
\hline
Region Surveyed & North Galactic Cap &  Bo\"{o}tes Field & Bo\"{o}tes Field & Lockman Hole\\
\hline
Field Area      & 10,000 deg$^2$ & 10 deg$^2$ & 250 deg$^2$ & 10 deg$^2$  \\
                & $b > 30^\circ$     &  $\alpha_{cen}=14^h32^m$ &$\alpha_{cen}=14^h32^m$ & $\alpha_{cen}=10^h49^m$  \\
                &                    &  $\delta_{cen} = 34^\circ16^\prime$ & $\delta_{cen} = 34^\circ16^\prime$ & $\delta_{cen} = 58^\circ20^\prime$ \\
Number of observations & 2           &  100 & 2 & 100 \\
Cadence         & 1 year & 1 day & 90 days & 1 day \\
Time per pointing & 2 min & 2 min & 2 min & 2 min \\
Detection threshold per epoch & 5 mJy & 5 mJy & 5 mJy & 5 mJy \\
Final detection threshold & 4 mJy & 1 mJy & 4 mJy & 1 mJy \\
Pointing angular spacing & $0.5^\circ$ &$0.5^\circ$ &$0.5^\circ$    &$0.5^\circ$    \\
Angular resolution & $100^{\prime\prime}$ & $100^{\prime\prime}$ & $100^{\prime\prime}$ & $100^{\prime\prime}$ \\
Position accuracy at $10\sigma$ & $\lsim 10^{\prime\prime}$& $\lsim 10^{\prime\prime}$& $\lsim 10^{\prime\prime}$ & $\lsim 10^{\prime\prime}$ \\
\enddata
\end{deluxetable}

\begin{deluxetable}{lrrrrrrrrr}
\tablecaption{Bo\"{o}tes Surveys\label{tab:bootes}}
\tablehead{\colhead{Survey} & \colhead{Frequency} & \colhead{Min. Flux } & \colhead{Resolution} & \colhead{FoV} \\
                            & \colhead{(GHz)}     &  \colhead{(mJy)}     & \colhead{arcsec $\times$ arcsec} & \colhead{deg$^2$}
}     
\startdata
WSRT-1400 & 1.4 & 0.14 & $13 \times 27$ & 7.0 \\
NVSS & 1.4 & 2.5 & $45 \times 45$ & $3\times 10^4$ \\
FIRST & 1.4 & 1.0 & $5 \times 5$ & $10^4$ \\
PiGSS & 3.1 & 0.8 & $100 \times 100$ & 11.7 \\
GB6 & 4.9 & 18 & $130 \times 130$ & $2\times 10^4$ \\
\enddata
\end{deluxetable}

\begin{deluxetable}{lrrrrrrrrrrrr}
\rotate
\tabletypesize{\scriptsize}
\tablecaption{PiGSS Sources \label{tab:sources}}
\tablehead{\colhead{Name} & \colhead{RA} & \colhead{Dec} & \colhead{$\Delta\alpha$} & \colhead{$\Delta\delta$}
& \colhead{$b_{maj}$} & \colhead{$b_{min}$} & \colhead{$\phi$} 
& \colhead{$b_{maj}^\prime$} & \colhead{$b_{min}^\prime$} & \colhead{$\phi^\prime$} 
& \colhead{Flux} & \colhead{Err} \\
                          &  (hours)     &  (deg)        &  (arcsec)              & (arcsec) 
& (arcsec)            & (arcsec)            & (deg)            
& (arcsec)            & (arcsec)            & (deg)            
& (mJy)          & (mJy) }
\startdata
J142318+344210 & 14.38826 & 34.70267 &  17.9 &  17.8 & 100.0 & 100.0 &    0.0 &   0.0 &   0.0 &    0.0 &   18.01 &    4.45 \\
J142402+344518 & 14.40044 & 34.75503 &  11.4 &  11.3 & 150.4 & 122.6 &   55.0 & 112.3 &  70.9 &   88.7 &   16.49 &    2.59 \\
J142421+343857 & 14.40570 & 34.64913 &  19.5 &  19.5 & 100.0 & 100.0 &    0.0 &   0.0 &   0.0 &    0.0 &    6.21 &    1.68 \\
J142426+343602 & 14.40719 & 34.60060 &   3.5 &   3.5 & 136.8 & 117.5 &   65.8 &  93.4 &  61.7 &    9.9 &   30.33 &    1.47 \\
J142430+341914 & 14.40831 & 34.32064 &  19.8 &  19.8 & 100.0 & 100.0 &    0.0 &   0.0 &   0.0 &    0.0 &    4.04 &    1.11 \\
J142440+343757 & 14.41101 & 34.63258 &   9.8 &   9.5 & 185.4 & 103.9 &   63.9 & 156.1 &  28.2 &  -61.2 &   10.00 &    1.32 \\
J142445+341832 & 14.41253 & 34.30878 &   2.1 &   2.1 & 111.5 & 102.7 &  -74.7 &  49.3 &  23.4 &  -40.0 &   42.03 &    1.22 \\
J142447+345317 & 14.41295 & 34.88818 &  15.9 &  15.9 & 100.0 & 100.0 &    0.0 &   0.0 &   0.0 &    0.0 &    8.54 &    1.88 \\
J142448+340957 & 14.41339 & 34.16577 &  10.7 &  10.7 & 100.0 & 100.0 &    0.0 &   0.0 &   0.0 &    0.0 &    6.18 &    0.92 \\
J142458+342527 & 14.41625 & 34.42421 &  10.3 &  10.3 & 164.4 & 105.3 &   66.5 & 130.5 &  33.0 &  -30.2 &    5.89 &    0.84 \\
J142503+334405 & 14.41761 & 33.73465 &  18.4 &  18.4 & 100.0 & 100.0 &    0.0 &   0.0 &   0.0 &    0.0 &    4.90 &    1.24 \\
J142516+345310 & 14.42118 & 34.88600 &   2.5 &   2.5 & 117.6 & 105.3 &   88.9 &  61.9 &  33.0 &  -53.6 &   38.86 &    1.34 \\
J142517+341606 & 14.42129 & 34.26831 &   5.0 &   5.0 & 100.0 & 100.0 &    0.0 &   0.0 &   0.0 &    0.0 &    9.71 &    0.67 \\
J142523+340944 & 14.42302 & 34.16212 &   5.3 &   5.3 & 100.0 & 100.0 &    0.0 &   0.0 &   0.0 &    0.0 &    8.83 &    0.64 \\
J142536+331215 & 14.42679 & 33.20403 &  10.1 &  10.2 & 140.3 & 115.9 &  -26.5 &  98.4 &  58.6 &   78.3 &   16.39 &    2.29 \\
J142541+345848 & 14.42811 & 34.98005 &   1.7 &   1.7 & 130.0 & 110.2 &   67.6 &  83.1 &  46.3 &   86.8 &  105.85 &    2.45 \\
J142543+335544 & 14.42853 & 33.92887 &   1.6 &   1.6 & 100.0 & 100.0 &    0.0 &   0.0 &   0.0 &    0.0 &  123.69 &    2.68 \\
J142558+351311 & 14.43279 & 35.21974 &  18.8 &  18.8 & 100.0 & 100.0 &    0.0 &   0.0 &   0.0 &    0.0 &    5.43 &    1.41 \\
J142601+343129 & 14.43357 & 34.52459 &  17.7 &  17.7 & 100.0 & 100.0 &    0.0 &   0.0 &   0.0 &    0.0 &    2.69 &    0.66 \\
J142607+340433 & 14.43531 & 34.07572 &   1.6 &   1.6 & 100.0 & 100.0 &    0.0 &   0.0 &   0.0 &    0.0 &   27.45 &    0.61 \\

\enddata
\tablecomments{First 20 lines of the table; full table is included as data file.}
\end{deluxetable}

\begin{deluxetable}{lrrrrrrrrrr}
\rotate
\tabletypesize{\scriptsize}
\tablecaption{Radio Matches to PiGSS Sources \label{tab:matches}}
\tablehead{\colhead{Name} & \colhead{$S_{1400}$} & \colhead{$\alpha_{1400}$} & \colhead{$\Delta\alpha_{1400}$} &\colhead{$S_{NVSS}$} & \colhead{$\alpha_{NVSS}$} & \colhead{$\Delta\alpha_{NVSS}$} &\colhead{$S_{4850}$} & \colhead{$\alpha_{4850}$} & \colhead{$\Delta\alpha_{4850}$} \\
                          & \colhead{(mJy)}      &                           &                                 &\colhead{(mJy)}      &                           &                                 &\colhead{(mJy)}      &                           &                                 }
\startdata
J142318+344210 &    0.00 &  0.00 &  0.00 &   34.30 & -0.81 &  0.31 &    0.00 &  0.00 &  0.00 \\ 
J142426+343602 &    0.00 &  0.00 &  0.00 &   47.60 & -0.57 &  0.06 &    0.00 &  0.00 &  0.00 \\ 
J142440+343757 &    0.00 &  0.00 &  0.00 &   15.90 & -0.59 &  0.17 &    0.00 &  0.00 &  0.00 \\ 
J142445+341832 &    0.00 &  0.00 &  0.00 &   82.10 & -0.85 &  0.04 &   22.00 & -1.44 & -0.41 \\ 
J142447+345317 &    0.00 &  0.00 &  0.00 &   20.70 & -1.12 &  0.28 &    0.00 &  0.00 &  0.00 \\ 
J142448+340957 &    0.00 &  0.00 &  0.00 &    8.10 & -0.34 &  0.20 &    0.00 &  0.00 &  0.00 \\ 
J142516+345310 &    0.00 &  0.00 &  0.00 &   65.90 & -0.67 &  0.04 &   28.00 & -0.73 & -0.33 \\ 
J142517+341606 &    0.00 &  0.00 &  0.00 &   16.30 & -0.65 &  0.10 &    0.00 &  0.00 &  0.00 \\ 
J142523+340944 &    0.00 &  0.00 &  0.00 &   20.60 & -1.07 &  0.10 &    0.00 &  0.00 &  0.00 \\ 
J142541+345848 &  262.60 & -1.15 &  0.06 &  154.70 & -0.48 &  0.03 &   69.00 & -0.95 & -0.23 \\ 
J142543+335544 &  322.78 & -1.21 &  0.06 &  303.10 & -1.13 &  0.03 &   79.00 & -0.99 & -0.23 \\ 
J142558+351311 &    0.00 &  0.00 &  0.00 &    9.30 & -0.68 &  0.34 &    0.00 &  0.00 &  0.00 \\ 
J142601+343129 &    5.43 & -0.89 &  0.31 &    3.60 & -0.37 &  0.36 &    0.00 &  0.00 &  0.00 \\ 
J142607+340433 &   37.73 & -0.40 &  0.06 &   34.60 & -0.29 &  0.03 &   28.00 &  0.04 & -0.32 \\ 
J142609+333949 &   29.78 & -0.89 &  0.09 &   27.80 & -0.81 &  0.08 &    0.00 &  0.00 &  0.00 \\ 
J142620+335127 &   13.96 & -1.30 &  0.16 &   11.50 & -1.05 &  0.17 &    0.00 &  0.00 &  0.00 \\ 
J142621+344021 &   87.01 & -0.74 &  0.05 &   38.60 &  0.29 &  0.02 &   36.00 & -0.66 & -0.31 \\ 
J142621+340937 &    6.51 & -1.09 &  0.25 &    5.00 & -0.76 &  0.27 &    0.00 &  0.00 &  0.00 \\ 
J142623+334643 &    4.51 &  1.28 &  0.09 &    2.60 &  1.98 &  0.25 &    0.00 &  0.00 &  0.00 \\ 
J142632+350831 &    0.00 &  0.00 &  0.00 &   95.00 & -1.13 &  0.05 &   28.00 & -0.72 & -0.33 \\ 

\enddata
\tablecomments{Zeros indicate no match.  
First 20 lines of the table; full table is included as data file.}
\end{deluxetable}


\end{document}